\documentclass[aps,pra,twocolumn]{revtex4-2}

\usepackage{amsmath,amsfonts,qcircuit,bbm}
\usepackage[colorlinks=true]{hyperref}
\usepackage[dvipsnames]{xcolor}
\usepackage{graphicx}
\usepackage{multirow}

\newcommand{\crt}[1]{\hat{c}_{#1}^\dagger}
\newcommand{\dst}[1]{\hat{c}_{#1}^{\phantom{\dagger}}}
\newcommand{\device}[1]{{{\em{ibmq$\_$#1}}}}
\newcommand{\resub}[1]{{\color{black}#1}}

\begin{document}

\title{Quantum simulations of molecular systems with intrinsic atomic orbitals}

\author{Stefano Barison}
\affiliation{Universit\`a degli Studi di Milano, Dipartimento di Fisica ``Aldo Pontremoli'', via Celoria 16, I-20133 Milano, Italy}
\author{Davide E. Galli}
\affiliation{Universit\`a degli Studi di Milano, Dipartimento di Fisica ``Aldo Pontremoli'', via Celoria 16, I-20133 Milano, Italy}
\author{Mario Motta}
\affiliation{IBM Quantum, IBM Research Almaden, 650 Harry Road, San Jose, CA 95120, USA}

\begin{abstract}

Quantum simulations of molecular systems on quantum computers 
often employ minimal basis sets of Gaussian orbitals. In comparison with more realistic basis sets, 
quantum simulations employing minimal basis sets require fewer qubits and quantum gates, 
but yield results of lower accuracy.

A natural strategy to achieve more accurate results is to increase the basis set size,
which in turn requires increasing the number of qubits and quantum gates.
Here we explore the use of intrinsic atomic orbitals (IAOs) in quantum simulations 
of molecules, to improve the accuracy of energies and properties at the same 
computational cost required by a minimal basis.

\resub{We investigate ground-state energies and one- and two-body density
operators in the framework of the variational quantum eigensolver,
employing and comparing different Ans\"{a}tze.
We also demonstrate the use of this approach in the calculation of 
ground- and excited-states energies of small molecules by a combination of quantum algorithms, using IBM Quantum computers.}

\end{abstract}
\maketitle

\section{Introduction}

The simulation of quantum many-body systems has long been recognized
as an application for quantum computers \cite{feynman_1982,lloyd_1996,abrams_1997,abrams_1997,georgescu2014quantum,abrams_1997,cao2019quantum,mcardle2020quantum,bauer_2020}.
While contemporary quantum devices and algorithms have enabled the 
simulation of ground- and excited-state properties of a variety of
systems \cite{kandala2017hardware}, quantum computing is still an 
emerging technology with limited simulation capabilities.
In the field of quantum chemistry, the limitations of quantum 
devices, classical simulators and quantum algorithms have resulted
in most quantum electronic structure simulations reported to date 
employing minimal basis sets of Gaussian orbitals \cite{o2016scalable,kandala2017hardware,rice2020quantum} or active 
spaces constructed on the basis of preliminary correlated classical
simulations \cite{gao2019computational}.

Such simulations have profound theoretical interest and represent
a driving force in the development of quantum devices, simulators and
algorithms, but they are far from returning the high-accuracy results needed 
by the quantum simulation of molecules.

Achieving this goal typically requires increasing significantly the 
number of qubits and quantum gates, and implementing sophisticated 
techniques to increase the representation accuracy of qubits \cite{takeshita2020increasing,motta2020quantum}.
Techniques that can improve the accuracy of quantum simulations without 
extra quantum resources and without reliance on preliminary classical simulations 
thus become desirable.

In the present work, we explore the use of intrinsic atomic orbitals
\cite{knizia2013intrinsic,senjean2020generalization} (IAOs) in the quantum simulation of molecular systems.
IAOs define atomic core and valence orbitals, polarized by the molecular 
environment, which can exactly represent self-consistent field wave 
functions, through a remarkably simple algebraic construction
\cite{knizia2013intrinsic} free from input from correlated many-body calculations. 
IAOs yielded accurate evaluations of a variety of chemical properties
in different environments and supported the understanding of molecular properties and the development of computational techniques  \cite{knizia2013intrinsic,schwilkIAO,ManzIAO,WestIAO,ElviraIAO,SchneiderIAO}.

Using the bond cleavage of several small molecules as an application, we demonstrate the integration of IAOs in a variety of quantum algorithms, 
using classical simulators of quantum computers and IBM Quantum hardware. We discuss strengths and weaknesses of the migration from minimal bases to IAOs, 
and identify the perturbative treatment of dynamical correlation from non-valence virtual orbitals as a way to further improve quantum simulations based on IAOs.

\section{Methods}
\label{sec:methods}

\subsection{Intrinsic Atomic Orbitals}
\label{sec:iao}

The IAO construction aims at combining the best 
properties of a set of molecular orbitals (MOs) 
$| \chi_m \rangle = \sum_{a} C_{am} 
| \varphi_a \rangle$ computed at mean-field level 
in a large basis set $B_1 = \{ \varphi_a \}_a$, 
and of a valence basis $B_2 = \{ \tilde{\rho}_b \}_b$ of atomic orbitals (AOs).
Here, given a molecule with geometry $G = \{ (Z_k , {\bf{R}}_k ) \}_{k=1}^{N_A}$, 
where $Z_k$ are the atomic numbers and ${\bf{R}}_k$ the positions of the constituent atoms, 
we choose $B_1 = \cup_k B(Z_k;{\bf{R}}_k)$, where $B(Z_k;{\bf{R}}_k)$ is a set of Gaussian orbitals for atom $k$ (e.g. Dunning's correlation consistent bases with polarized and multiple valence orbitals, usually abbreviated in cc-pVxZ \cite{dunning1989gaussian}).
On the other hand, to construct $B_2$, for every atom in the molecule we perform a single-atom Hartree-Fock calculation with basis $B(Z_k;{\bf{R}}_k)$, yielding a set of core, valence and external orbitals for that particular atom, and we append the core and valence orbitals to the basis $B_2$.
Since a common set of single-atom bases $B(Z_k;{\bf{R}}_k)$ is used in the construction of both $B_1$ and $B_2$, then $B_2$ is a proper subset of $B_1$.

The MOs can of course reproduce the mean-field wavefunction from which they are defined, 
but cannot be clearly associated with any atom,
which complicates the interpretation of the wavefunction and of its properties.
The AOs, though naturally associated with an atom, give an inaccurate representation of the MOs, as they contain no polarization due to the molecular environment.
The IAO basis is \resub{then constructed} by forming a set of polarized AOs $\{ \rho_b \}_b$ that, at variance with the AOs in $B_2$, can exactly express occupied MOs $| \chi_i \rangle$. 
First, the projectors $P = \sum_i | \chi_i \rangle \langle \chi_i |$, $Q = \mathbbm{1} - P$ onto occupied and virtual MOs are defined.
This allows to define the projectors
\begin{equation}
\begin{split}
\label{eq:iao_prj}
P_{12} &= \sum_{\phi_{a},\phi_{b} \in B_1} S^{ab} |\phi_a\rangle \langle \phi_{b}| \quad, \\
P_{21} &= \sum_{\tilde{\rho}_{c},\tilde{\rho}_{d} \in B_2} \tilde{S}^{cd} |\tilde{\rho}_c\rangle \langle \tilde{\rho}_{d}| \quad, \\
\end{split}
\end{equation}
onto the bases $B_1$ and $B_2$, where $S^{ab}$ and $\tilde{S}^{cd}$ are the inverse overlap matrices in $B_1$ and $B_2$ respectively.
Then, a set of depolarized occupied MOs $| \tilde{\chi}_i \rangle = P_{12} P_{21} | \chi_i \rangle$ is constructed by projecting the original, polarized, occupied MOs onto the
AO basis $B_2$ and immersing the projected MOs in the original basis $B_1$. The depolarized occupied MOs are used to define the projectors $\tilde{P} = \sum_i | \tilde{\chi}_i \rangle \langle \tilde{\chi}_i |$ and $\tilde{Q} = \mathbbm{1} - \tilde{P}$, and the IAOs are obtained as
\begin{equation}
\label{eq:iao}
| \rho_b \rangle = 
\left( P \tilde{P} + Q \tilde{Q} \right)
P_{12} | \tilde{\rho}_b \rangle \quad .
\end{equation}
\resub{Therefore, IAOs are constructed through a sequence of simple and natural algebraic operations.
In addition to the projection \eqref{eq:iao}, we orthonormalize the IAO basis to ensure the satisfaction of canonical anticommutation relations between second-quantization operators, 
and we perform a Foster-Boys localization of the IAOs to enhance their spatial locality \cite{foster1960canonical}}.

Finally, we consider the Born-Oppenheimer approximation of the molecular Hamiltonian \cite{born_oppenheimer1927}

\begin{equation}
\label{eq:es_ham}
H = E_0 + \sum_{ \substack{pq\\\sigma} } h_{pq} \, \crt{p\sigma}
\dst{q\sigma}
+ 
\sum_{ \substack{prqs\\\sigma\tau} }
\frac{(pr|qs)}{2} \, 
\hat{c}_{p\sigma}^\dagger 
\hat{c}_{q\tau}^\dagger 
\hat{c}_{s\tau}^{\phantom{\dagger}}
\hat{c}_{r\sigma}^{\phantom{\dagger}} \, ,
\end{equation}

where $E_0$ indicates the repulsion between nuclei of the molecule, $h_{pq}$ is the one-body part of the Hamiltonian, containing kinetic energy of the electron plus the interaction with the fixed nuclei and $(pr|qs)$ is the electron-electron repulsion integral.
Once the IAOs are defined, we fold the Hamiltonian in Eq. \ref{eq:es_ham} through a standard atomic orbitals to molecular orbitals (ao2mo) transformation, from the $B_1$ to the orthonormalized IAO basis. 
In this work, we relied on the frozen-core approximation, since the basis sets we employed lack core-valence correlation effects.

\subsection{Ground- and excited-states algorithms}
\label{sec:computational}

We explored the ground and excited states of the Hamiltonian (\ref{eq:es_ham})
with several techniques.
Here, we focused on the the variational quantum eigensolver  \cite{peruzzo2014variational,mcclean2016theory}
and quantum imaginary-time evolution \cite{motta_2019} methods for ground-state studies.
In the Appendix \ref{app:qeom}, we also investigate the quantum equation-of-motion \cite{ollitrault2019quantum} method for excited-state studies.

\subsubsection{Variational Quantum Eigensolver}
\label{sec:vqe}

Variational quantum state preparation algorithms are widely used on contemporary 
quantum devices. These algorithms define a set of Ansatz states approximating the 
ground state of a target Hamiltonian, of the form 
$| \Psi(\theta) \rangle = \hat{U}(\theta) | \Psi_0 \rangle$, 
$\theta \in \Theta \subseteq \mathbb{R}^n$.
In other words, a parametrized quantum circuit
$\hat{U}(\theta)$ is applied to an initial wavefunction $| \Psi_0 \rangle$. The best approximation to the ground state in the set of Ansatz states is found by minimizing the energy $E(\theta) = \langle \Psi(\theta) | \hat{H} | \Psi(\theta) \rangle$ as a function of the parameters $\theta$ using a classical optimization algorithm \cite{peruzzo2014variational,mcclean2016theory}.
This algorithmic workflow, termed variational quantum eigensolver (VQE) \cite{peruzzo2014variational} in the quantum simulation literature, is a heuristic technique for ground-state approximation. 
Its accuracy and computational cost are determined by the form of the circuit $\hat{U}(\theta)$.

Within VQE, we compare different Ans\"{a}tze $\hat{U}(\theta)$ :
\begin{enumerate}
\item the quantum unitary coupled cluster with single and double excitations (q-UCCSD) , where $\hat{U}(\theta)$ 
is a qubit representation of the operator
\cite{kutzelnigg1982quantum,kutzelnigg1983quantum,kutzelnigg1985quantum,barkoutsos2018quantum}
\begin{equation}
\begin{split}
&\hat{U}_{\mathrm{q-UCCSD}}(\theta) = e^{ \hat{T} - \hat{T}^\dagger } \quad, \\
&\hat{T} = \sum_{ai} \theta^a_i \, \crt{a} \dst{i} + \sum_{abij} \theta^{ab}_{ij} \, \crt{a} \crt{b} \dst{j} \dst{i} \quad,
\end{split}
\end{equation}
with $ij$ occupied and $ab$ virtual in the mean-field reference state.
The q-UCCSD quantum circuit is given in Ref. \cite{barkoutsos2018quantum}.
\item the hardware-efficient $R_y$ Ansatz with linear connectivity \cite{kandala2017hardware} which, for a register of $n$ qubits and an 
Ansatz of depth $d$, takes the form
\begin{equation}
\hat{U}_{R_y}(\theta) = 
\left[ \prod_{i=0}^{n-1} \hat{R}^{(i)}_y(\theta^{d}_{i}) \right]
\prod_{\ell=0}^{d-1} 
\Bigg[ \prod_{i=0}^{n-2} U^{\text{ent}}_{i,i+1}  \prod_{i=0}^{n-1} \hat{R}^{(i)}_y(\theta^{\ell}_{i}) \Bigg] \quad,
\end{equation}
where $U^{\text{ent}}_{i,i+1}$ is an entangler gate of choice, in this case a $\mathsf{CNOT}$ gate with control qubit $i$ and target qubit $i+1$, and $\hat{R}^{(i)}_y(\theta_{i}^{\ell})$ is the $\ell$-th $Y$-rotation of qubit $i$ by an angle $\theta_{i}^{\ell}$ 
\item the hardware-efficient SO(4) Ansatz, of the form
\begin{equation}
\label{eq:so4}
\hat{U}_{\mathrm{SO}(4)}(\theta) = \prod_{\ell=0}^{d-1} 
\Bigg[ \prod_{(ij) \in N} \hat{u}_{ij}(\theta_{ij}^{\ell}) \Bigg] \quad,
\end{equation}
where $u_{ij}$ is a two-qubit gate in the $\mathrm{SO}(4)$ group. 
It is well-known \cite{vatan2004optimal} that a two-qubit gate in the $\mathrm{SO}(4)$ group can be written (as shown in Fig. \ref{fig:figure_3}) as a product of 2 Hadamard, 4 S, 2 $\mathsf{CNOT}$, and two single-qubit $u_3$ gates. Single-qubit $u_3$ gates are defined as
\begin{equation}
u_3(\theta,\phi,\lambda) = R_z(\phi) R_x\left(-\frac{\pi}{2}\right) R_z(\theta) R_x\left(\frac{\pi}{2}\right) R_z(\lambda)
\end{equation}
where $\theta,\phi,\lambda$ are three angles and $R_{x,y,z}$ are single-qubit $X$, $Y$, and $Z$ rotations respectively.

\end{enumerate}

\subsubsection{Quantum imaginary-time evolution}
\label{sec:qite}

Quantum imaginary-time evolution (QITE) \cite{motta_2019,yeter-aydeniz_practical_2020,yeter-aydeniz_scattering_2020,nishi_2020,gomes_2020} is an alternative and complementary technique to VQE and other heuristic quantum algorithms for ground-state search.
QITE is an Ansatz-independent technique, that approaches the ground state of a quantum system by applying the following imaginary-time evolution (ITE) map on a trial wavefunction $| \Psi_T \rangle$,
\begin{equation}
| \Psi_\beta \rangle 
= 
\frac
{e^{-\beta \hat{H}} | \Psi_T \rangle}
{\| e^{-\beta \hat{H}} \Psi_T \|}
\quad.
\end{equation}
The ITE is divided in a large number $n_\beta$ of steps of length $\Delta \tau = \beta/n_\beta$ and ITE under a single step is approximated by a Trotter decomposition,
\begin{equation}
e^{-\beta \hat{H}} \simeq \prod_m e^{-\beta \hat{h}[m]} \quad ,
\end{equation}
where $\hat{H} = \sum_m \hat{h}[m]$ is a representation of the Hamiltonian as a sum of local operators. ITE under a single imaginary-time step and a single local term of the Hamiltonian is approximated by a unitary transformation, that is equal to the exponential of a linear combinations of local operators $P_\mu$,
\begin{equation}
\frac
{e^{-\Delta\tau \hat{h}[\mu]} | \Psi \rangle}
{\| e^{-\Delta\tau \hat{h}[\mu]} \Psi \|}
\simeq
e^{ i \sum_\mu x_\mu P_\mu } | \Psi \rangle \quad .
\end{equation}
The coefficients $x_\mu$ are determined \cite{motta_2019} solving
a linear system of the form $A x = b$, with
\begin{equation}
A_{\mu\nu} = \langle \Psi | P_\mu P_\nu | \Psi \rangle
\quad,\quad
b_\mu = \langle \Psi | P_\mu \hat{h}[m] | \Psi \rangle
\quad.
\end{equation}
The QITE simulations reported in this work
are carried out in a two-orbital space.
For such a problem, additional simplifications are possible, which are
listed and discussed in the Appendix \ref{app:qite}.

\subsection{Evaluation of density matrices}
\label{sec:rdm}

Once the optimal state $| \Psi \rangle$ is found, ground-state properties can be computed as expectation values of suitable 
qubit operators. Here we consider the case of one- and two-body density matrices,
\begin{equation}
\label{eq:rdms}
\begin{split}
\rho^{(\sigma)}_{pr} &= \langle \Psi | \crt{p\sigma} \dst{r\sigma} | \Psi \rangle
\;, \\
\rho^{(\sigma,\tau)}_{prqs} &= \langle \Psi | \crt{p\sigma} \crt{q\tau} \dst{s\tau} \dst{r\sigma} | \Psi \rangle
\;, \\
\end{split}
\end{equation}
which are useful for a variety of applications, from computing correlation functions 
to understanding electron entanglement and molecular bonding \cite{lowdin1,lowdin2,lowdin3}
and performing orbital relaxation \cite{werner1985second,head1988optimization,sherrill1998energies}.

The operators \eqref{eq:rdms} can be mapped onto qubit operators using standard techniques. 
For example, in the Jordan-Wigner \cite{jordan1993paulische,bravyi2002fermionic,seeley2012bravyi}
representation,
\begin{equation}
\crt{p\sigma} = 
\left\{
\begin{array}{ll}
(S_+)_p \sigma^{z}_{p-1} \dots  \sigma^{z}_0 & \sigma = \, \uparrow \\
(S_+)_{n+p} \sigma^{z}_{n+p-1} \dots \sigma^{z}_0 & \sigma = \, \downarrow \\
\end{array}
\right.
\end{equation}
where
\begin{equation}
S_+ = \frac{\sigma^{x}+i\sigma^{y}}{2} \,\,\, \text{and} \,\,\, S_- = \frac{\sigma^{x}-i\sigma^{y}}{2} \, ,
\end{equation}
$n$ is the size of the IAO basis and $\sigma^{\mu}$ with $\mu \in \{x,y,z\}$ are the standard Pauli $x,y$ and $z$ operators, respectively.
Therefore
\begin{equation}
\rho^{(\sigma)} = \langle \Psi | X^\sigma_{pr} | \Psi \rangle
\;,
\end{equation}
with
\begin{equation}
X^\uparrow_{pr} =
\left\{
\begin{array}{ll}
(S_+)_p \sigma^{z}_{p-1} \dots \sigma^{z}_{r+1} (S_-)_r & \text{if} \quad p>r \\
\frac{1-\sigma^{z}_p}{2} & \text{if} \quad p=r \\
(S_-)_r \sigma^{z}_{r-1} \dots \sigma^{z}_{p+1} (S_+)_p & \text{if} \quad p<r \\
\end{array}
\right.
\end{equation}
and
\begin{equation}
X^\downarrow_{pr} =
\left\{
\begin{array}{ll}
(S_+)_{p+n} \sigma^{z}_{p+n-1} \dots \sigma^{z}_{r+n+1} (S_-)_r & \text{if} \quad p>r \\
\frac{1-\sigma^{z}_{n+p}}{2} & \text{if} \quad p=r \\
(S_-)_{r+n} \sigma^{z}_{r-1} \dots \sigma^{z}_{p+n+1} (S_+)_{p+n} & \text{if} \quad p<r \\
\end{array}
\right.
\end{equation}
In a similar way,
\begin{equation}
\rho^{(\sigma\tau)}_{prqs} = \langle \Psi | X^\sigma_{pr} X^\tau_{qs} | \Psi \rangle - \delta_{qr} \delta_{\sigma\tau} \langle \Psi | X^\sigma_{ps} | \Psi \rangle \quad .
\end{equation}

\subsection{Variational quantum subspace expansion}

Incorporating dynamical correlation from non-valence virtual orbitals is important to improve the quantitative accuracy of simulations based on IAOs:
here, we demonstrate how to partly overcome this limitation, using a simplified implementation of the virtual quantum subspace expansion technique (VQSE).
This technique, proposed by Takeshita et al \cite{takeshita2020increasing}, 
introduces contributions from virtual orbitals lying outside a chosen active space in a systematic way.
The starting point of VQSE is a reference function $\Psi_0$ constructed in a set of active orbitals from a large basis.
Here, active-space orbitals are linear combinations of IAOs, denoted with lowercase letters, $p \in A$.
Uppercase letters $P \in B_1$ denote orthonormal orbitals in the basis used to construct IAOs.

Next, VQSE introduces a set of expansion operators. Here, we choose
\begin{equation}
\begin{split}
| \Psi \rangle 
&= \left[ \alpha + \beta_{P r} \crt{P \sigma} \dst{r \sigma} + \gamma_{TuVw} \crt{T \sigma} \crt{V \tau} \dst{w \tau} \dst{u \sigma} \right]  | \Psi_0 \rangle \\
&= \left[ \alpha + \beta_{P r} E_{Pr} + \gamma_{TuVw} E_{TuVw} \right] | \Psi_0 \rangle \quad .\\
\end{split}
\end{equation}
Electrons are excited from active to generic orbitals, excitation operators are summed over spin polarizations $\sigma,\tau$
and Einstein's summation convention is used.
Note that the reference wavefunction has no components outside the active space $A$, 
and therefore contraction over orbitals outside $A$ can be computed analytically using Wick's theorem.

The amplitudes $ v = \left( \, \alpha \,\,\, \beta \,\,\, \gamma \, \right)^T$ are real-valued, and determined by solving a generalized eigenvalue equation $H v = E S v$.

Detailed calculation to obtain the explicit form of $H$ and $S$ can be found in Appendix \ref{app:vqse}.
The matrix elements of $H$ and $S$  are evaluated using data from a quantum device and subsequently diagonalized on a classical computer, to extract the lowest eigenvalue.
Although we relied on full diagonalization and extraction of the lowest eigenvalue for simplicity, a better scaling with basis size could easily be achieved using Davidson's algorithm.

In the present work, we focused on two-electron problems, where the the explicit form of $H$ and $S$ are defined by the active-space one- and two-body density matrices  \cite{takeshita2020increasing}, that we introduced in Section \ref{sec:rdm}.

\subsection{Software for classical and quantum simulations}
\label{sec:method_soft}

The calculations performed here involved initial 
pre-processing using the PySCF quantum chemistry
package \cite{sun2018pyscf,sun2020recent}.
PySCF was used to generate optimized mean-field 
states, Hamiltonian matrix elements in the IAO 
basis, and a reaction path for the NH$_3 \to$ NH$_2$ + H reaction by a collection of constrained geometry optimizations performed using Moller-Plesset perturbation theory \cite{moller_plesset1934} in a correlation consistent Dunning's basis, augmented with an extra diffuse function in each orbital angular momentum (MP2/aug-cc-pVTZ).
The restricted closed-shell Hartree-Fock (RHF) singlet state was  chosen as the initial state for all of the calculations  described here. 
Intrinsic atomic orbitals are computed as detailed in Section \ref{sec:iao}  and IAOs obtained from an underlying basis B are denoted as IAO/B.

Having selected a set of single-electron orbitals for each of the studied species, quantum computations were performed with quantum simulators and hardware.
We used IBM's open-source library for quantum computing, Qiskit \cite{aleksandrowicz2019qiskit}.
In particular, the library contains implementations of techniques to map the fermionic Fock space onto the Hilbert space of a register of qubits, and implementations of VQE and quantum equation-of-motion. 
In addition, a module for QITE simulations was composed using Qiskit subroutines. 
We use the tapering-off technique \cite{bravyi2017tapering,setia2019reducing} to account for molecular point group symmetries and reduce the number of qubits required for a simulation whenever possible.

In VQE simulations, we minimized the expectation value  of the Hamiltonian with respect to the parameters in the circuit. 
On simulators, optimizations were carried out using the L-BFGS-B and CG methods \cite{zhu1997algorithm,morales2011remark}, using the statevector simulator of Qiskit. 
On quantum hardware, optimizations were carried out using the gradient descent optimization method described in Appendix \ref{app:opt}. 
We performed quantum computations on quantum hardware using various 5-qubit IBM Quantum devices, specifically \device{rome}, \device{vigo} and \device{london}.

\subsection{Error mitigation techniques}
\label{sec:err_mitigation}
In order to improve the quality of noisy hardware experiments, we referred to readout error mitigation techniques included in Qiskit \cite{temme2017error}.

In particular, we used measurement calibration  to mitigate measurement errors.
Given a system of $N$ qubits, all $2^N$ basis input states are prepared and the probability of measuring counts in the other basis states is computed.
From these results, a calibration matrix is created and used to improve the results of subsequent experiments.

All the experiments proposed in this work required 2 qubits and no more than 2 CNOTs, indicating that the dominant source of noise was measurement error.
Running the 4 calibration circuits was sufficient to obtain good quality results.

Computing the calibration matrix becomes quickly unfeasible as the number of qubits increases, for this reason more efficient methods have been proposed \cite{nation_2022}.
When the depth of the circuit is increased, gates error will play a significant role on the quality of the results. 
In this case, gate error mitigation techniques can be adopted, such as zero noise or Richardson extrapolation \cite{temme2017error, Li_2017, Kandala_2019,Carbone_2022} or probabilistic error cancellation \cite{Berg_22}.

\section{Results}
\label{sec:results}

\subsection{Comparison between minimal bases and IAO}

\begin{table*}[t]
\centering\begin{tabular}{c| c |cc| cc | cc | cc | cc}
\hline\hline
\multicolumn{2}{c}{} & \multicolumn{2}{c}{H$_2$}  & \multicolumn{2}{c}{HeH$^+$} & \multicolumn{2}{c}{LiH} & \multicolumn{2}{c}{H$_2$O} & \multicolumn{2}{c}{NH$_3$}\\
\hline
basis  & method                  & $\Delta E$ [$\mathrm{E_h}$] & $R_{eq} [\mathrm{\AA}]$ & $\Delta E$ [$\mathrm{E_h}$] & $R_{eq} [\mathrm{\AA}]$ & $\Delta E$ [$\mathrm{E_h}$] & $R_{eq} [\mathrm{\AA}]$ & $\Delta E$ [$\mathrm{E_h}$] & $R_{eq} [\mathrm{\AA}]$ & $\Delta E$ [$\mathrm{E_h}$] & $R_{eq} [\mathrm{\AA}]$ \\
\hline
\multirow{5}{*}{STO-6G} & HF    & N/A        & 0.695(9) & N/A       &  0.937(9)  & N/A        & 1.482(5) & N/A        & 0.993(1) & N/A        & 1.024(2) \\
 & $R_y$, $d=1$                 & 0.2084(9)  & 0.715(7) & 0.0513(3) &  0.919(1)  & 0.0833(7)  & 1.482(7) & 0.1353(1)  & 0.970(7) & 0.1525(7)  & 1.024(3) \\
 & $\mathrm{SO}(4)$, $d=1$      & 0.2083(9)  & 0.715(7) & 0.0513(3) &  0.919(1)  & 0.0837(6)  & 1.482(8) & 0.1358(4)  & 0.941(5) & 0.1510(6)  & 1.024(2) \\
 & q-UCCSD                      & 0.2083(9)  & 0.715(7) & 0.0513(3) &  0.919(1)  & 0.1079(9)  & 1.522(1) & 0.1625(8)  & 1.006(5) & 0.1694(4)  & 1.058(2) \\
 & FCI                          & 0.2092(2)  & 0.715(7) & 0.0512(7) &  0.919(4)  & 0.1075(5)  & 1.522(2) &  0.1626(0) & 1.006(7) & 0.1694(4)  & 1.058(2) \\
\hline
\multirow{5}{*}{IAO} & HF      & N/A        & 0.716(4)  & N/A       &  0.770(4) & N/A        & 1.586(5) & N/A        & 0.949(7)   & N/A        & 0.995(1)  \\
& $R_y$, $d=1$                & 0.1721(7)  & 0.729(9)  & 0.0822(9) &  0.768(4) & 0.10589    & 1.586(6) & 0.1755(7)  & 0.921(9)   & 0.1774(9)  & 0.996(3)  \\
 & $\mathrm{SO}(4)$, $d=1$     & 0.1721(7)  & 0.729(9)  & 0.0822(9) &  0.768(4) & 0.1000(2)  & 1.584(2) & 0.1861(8)  & 0.924(8)   & 0.1760(6)  & 0.996(1)  \\
 & q-UCCSD                     & 0.1760(5)  & 0.729(9)  & 0.0822(9) &  0.768(4) & 0.1362(0)  & 1.612(9) & 0.1978(2)  & 0.965(6)   & 0.1849(6)  & 1.022(1)  \\
 & FCI                         & 0.1721(7)  & 0.729(9)  & 0.0822(6) &  0.767(8) & 0.1362(0)  & 1.612(9) & 0.1924(2)  & 0.965(8)   & 0.1923(4)  & 1.023(1)  \\
 \hline
  aug-cc-pVQZ & CCSD                        & 0.1798(4)  & 0.719(7)  & 0.0752(3) &  0.774(6) & 0.1388(2)  & 1.572(9) & 0.2309(1)  & 0.959(8)   & 0.2139(9)  & 1.007(8)  \\
\hline\hline
\end{tabular}
\caption{\resub{Dissociation energy and equilibrium bondlength  for different molecules, using various VQE Ans\"{a}tze at STO-6G and IAO/aug-cc-pVQZ level of theory, and from CCSD at aug-cc-pVQZ level of theory.
From here on, except when explicitly stated, the energies are reported in Hartree units ($\mathrm{E_h}$) and atomic distances in Angstrom units ($\mathrm{\AA}$).}
}
\label{tab:table_1}
\end{table*}

The migration from minimal to IAO bases, in quantum and classical simulations of molecules, has benefits and limitations.
On the one hand, use of IAOs reduces basis set errors at mean-field level, because IAOs are designed to reproduce mean-field results.
While basis set errors still affect chemical properties, and particularly correlation energies and response functions, 
their removal at mean-field level can improve the accuracy of many computational predictions, especially in chemical species that are sensitive to the presence of polarized and diffuse functions.
Furthermore, IAOs are based on a computationally inexpensive and general-purpose procedure, that enables accurate calculations of a variety of chemical properties \cite{knizia2013intrinsic,schwilkIAO,ManzIAO,WestIAO,ElviraIAO,SchneiderIAO}, and does not resort to preliminary correlated many-body calculations, e.g. of MP2 or complete active space self consistent field (CASSCF) type, which need to be converged and carefully designed to avoid biasing chemical properties \cite{shao2006advances,baader2006chemistry,malmqvist2008restricted,stein2016automated,ElviraIAO}.

The main limitation stemming from the use of IAOs is the presence of residual basis set errors, which can only be removed by simulating orbitals beyond the IAO basis, or with additional post-processing. 
However, unlike minimal bases, IAO bases are naturally embedded into larger basis of one-electron orbitals, because they are constructed from such a basis. 
As a result, IAOs can be employed to capture static electronic correlation in a valence space, while dynamical correlation originating from electronic transitions to orbitals outside the IAO space can be treated perturbatively, as in classical CASPT2 \cite{roos1982simple,andersson1990second,andersson1992second} and NEVPT2 calculations \cite{angeli2001introduction,angeli2001n,angeli2002n}, or in recently-proposed quantum-computing methods like VQSE \cite{takeshita2020increasing}.
In this sense, the migration from minimal to IAO bases can constitute an opportunity to integrate techniques to perturbatively capture dynamical correlation in the workflow of quantum simulations, as well as to compare, demonstrate, and develop such techniques.

To illustrate the difference between minimal and IAO bases, in Table \ref{tab:table_1} we study the dissociation of a single H atom from a few molecules, namely H$_2$, HeH$^+$, LiH, H$_2$O, and NH$_3$.
We evaluate the ground-state energy along the dissociation path at minimal basis set STO-6G and IAO/aug-cc-pVQZ level, using RHF and VQE with $R_y$, SO(4) and q-UCCSD Ans\"{a}tze.
As an approximation to the complete basis set limit, we perform a coupled cluster calculation with single and double excitations (CCSD) in the aug-cc-pVQZ basis (CCSD/aug-cc-pVQZ) \cite{cizek1966} .
Table \ref{tab:table_1} reports equilibrium bondlengths $R_{eq}$ and binding energies $\Delta E$.

\begin{figure}[b!]
\includegraphics[width=1.0\columnwidth]{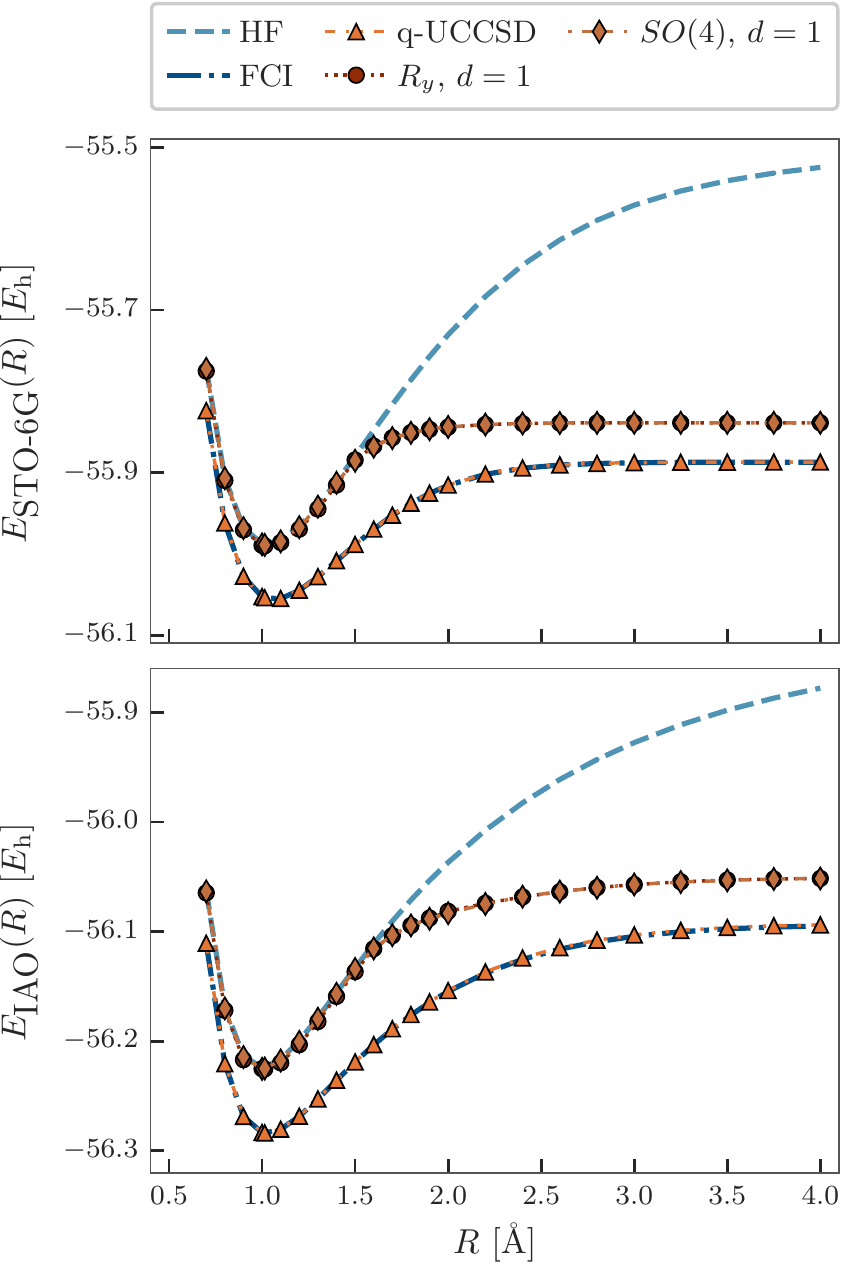} 
\caption{Ground-state potential energy curve of NH$_3$ along the NH$_3 \to$ NH$_2$ + H reaction path, at STO-6G (top) and IAO/aug-cc-pVQZ (bottom) level, using RHF (dashed light blue line), FCI (dark blue dash-dotted line) and VQE with $R_y$ (red circles), $SO(4)$ (dark orange diamonds) and q-UCCSD Ans\"{a}tze (orange triangles). $d$ indicates the depth of the Ansatz.
}
\label{fig:figure_1}
\end{figure}

For all studied species, FCI/IAO and q-UCCSD/IAO binding energies are in better agreement with CCSD/aug-cc-pVQZ binding energies than their counterparts at STO-6G level.
In particular, the mean absolute deviation $| \delta E(\textrm{FCI/IAO}) - \delta E(\textrm{CCSD/aug-cc-pVQZ}) |$
between FCI/IAO and FCI/STO-6G binding energies is 0.015(1) Hartree, whereas for FCI/STO-6G binding energies it is 0.039(1) Hartree. 
The improvement is more modest for hardware-efficient Ans\"{a}tze, which is not unexpected, in view of their heuristic nature.
Correspondingly, the mean absolute deviation $| R_{eq}(\textrm{FCI/IAO}) - R_{eq}(\textrm{CCSD/aug-cc-pVQZ}) |$ between FCI/IAO and FCI/STO-6G binding energies is 0.016(1) $\mathrm{\AA}$, whereas for FCI/STO-6G equilibrium bondlengths it is 0.059(1) $\mathrm{\AA}$. 

In Fig.~\ref{fig:figure_1}, we report the complete ground-state potential energy curve along the NH$_3 \to$ NH$_2$ + H reaction path.
We observe that VQE/q-UCCSD provides results of FCI-like accuracy, whereas $R_y$ and SO(4), though describing in a qualitatively correct way the dissociation limit, produce results of lower quality at lower computational cost.
The deviation between hardware-efficient Ansatze and VQE/q-UCCSD is maximal around the valley-ridge inflexion point $R \simeq 1.75 \mathrm{\AA}$, where the wavefunction has maximally multireference character.

\subsection{Hardware experiments}

\begin{figure}[h!]
\includegraphics[width=1.0\columnwidth]{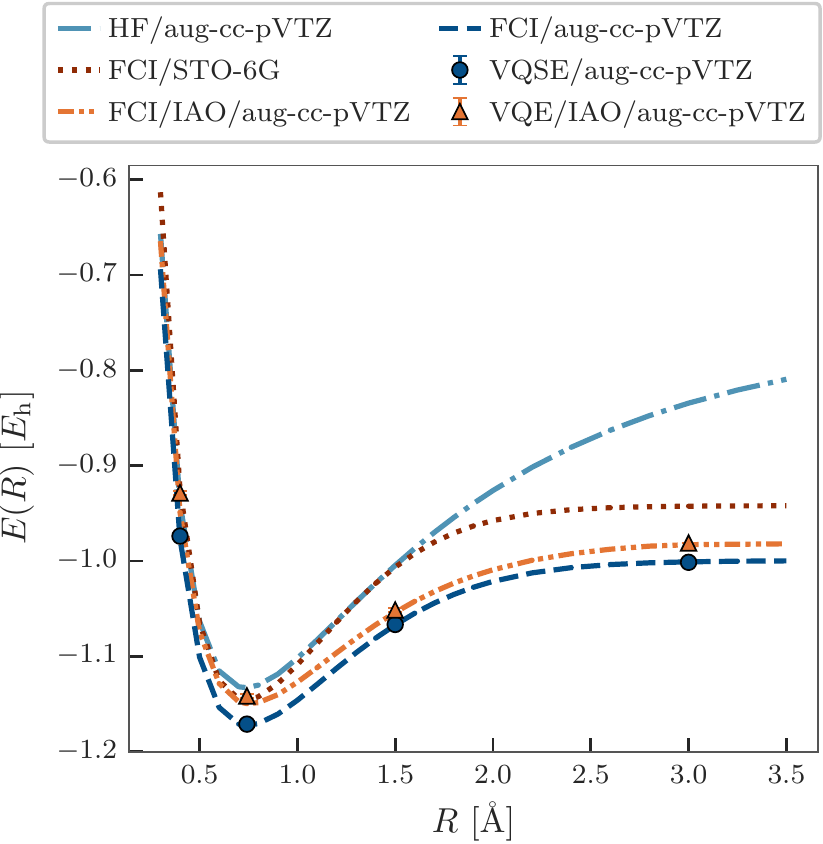}
\caption{Ground state potential energy surface of H$_2$ calculated using VQE and FCI with an IAO/aug-cc-pVTZ basis (orange triangles and dashdot-dotted lines), and VQSE and FCI with full aug-cc-pVTZ basis (dark blue circles and dashed line).
Statistical uncertainties on the lowest eigenvalue were obtained by sampling the active-space density matrices 10 times,repeating the embedding, contraction, and diagonalization procedures for each sample,and collecting statistics with standard procedures.}
\label{fig:figure_2}
\end{figure}

In this Subsection, we describe hardware experiments. As an illustrative application, in Fig.~\ref{fig:figure_2}, the potential energy surface of H$_2$ is computed using VQE with IAO/aug-cc-pVTZ basis.
Simulations required two qubits from the \device{rome} device, and employed an R$_y$ Ansatz with depth $d=1$.
Given the simplicity of this application, VQE results are statistically compatible with FCI results obtained at IAO/aug-cc-pVTZ level (agreement between orange triangles and line). 
As discussed in the previous section, improve the prediction of binding energies and equilibrium bondlengths over STO-6G (dotted line).
The VQE/IAO/aug-cc-pVTZ results correspond to the simulation of the full valence space of H$_2$, but are not sufficient to recover aug-cc-pVTZ results, 
because virtual orbitals outside the IAO basis are not included in the simulation (deviation between orange dashdot-dotted and dark blue dashed curves). 
We illustrate how this limitation can be overcome by computing VQSE energies, which are statistically compatible with FCI/aug-cc-pVTZ energies (agreement between red points and line).
 
As a more interesting application, in the remainder of this Subsection, we present the dissociation of NH$_3$.
Studying the full 7-orbital IAO basis requires 14 qubits using a second-quantization encoding, which can be reduced to 11 or 7 using qubit-reduction techniques \cite{bravyi2017tapering,eddins2021doubling}.
For illustrative purposes, and in order to use a number of qubits and gates compatible with simulation on 5-qubit devices, we constructed an active space from the IAO/aug-cc-pVQZ basis.

Specifically, for every geometry along the reaction path, we performed an MP2 calculation in the IAO/aug-cc-pVQZ basis.
Structures were relaxed in the estimation of the binding energies.
We constructed an active space using the highest-unoccupied and the lowest-unoccupied natural orbitals (HONO/LUNO active space). 

\begin{figure}[h!]
\includegraphics[width=1.0\columnwidth]{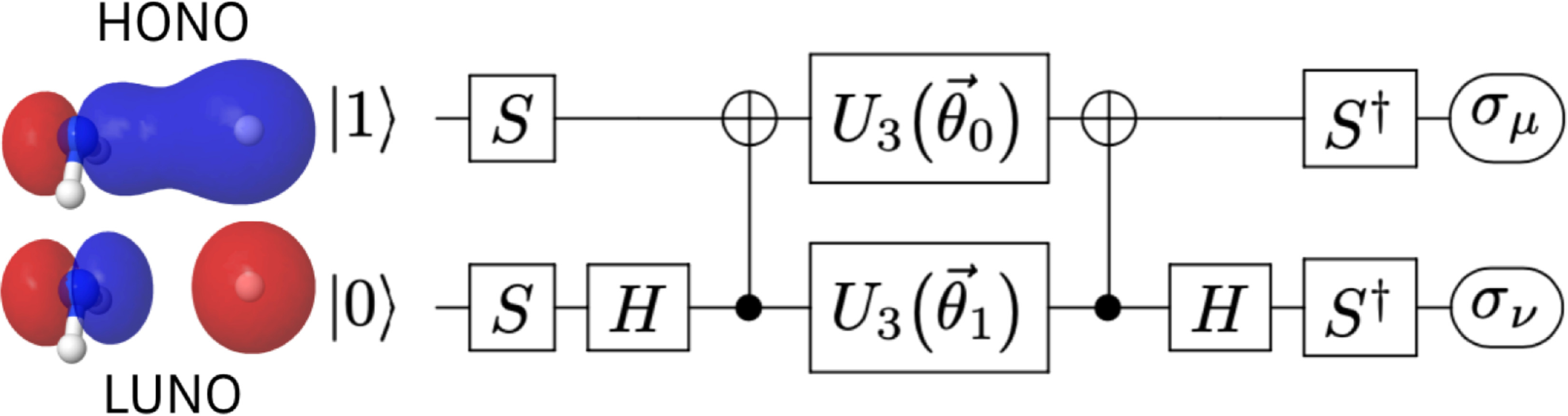} 
\caption{Left: highest-occupied (HONO, top) and lowest unoccupied (LUNO, bottom) natural orbitals from an MP2 calculation using the IAO/aug-cc-pVQZ basis.
Right: quantum circuit for the VQE calculations with SO(4) Ansatz in the HONO/LUNO subspace. $U_3$ denotes an SU(2) gate, parametrized by 3 Euler angles.
$\sigma^\mu$ and $\sigma^\nu$ are Pauli operators appearing in the qubit representation of the active space Hamiltonian,
$\hat{H} = \sum_{\mu\nu=0}^3 \eta_{\mu\nu} \, \sigma^\mu \otimes \sigma^\nu$.}
\label{fig:figure_3}
\end{figure}

The HONO and LUNO are linear combinations of the 1s-like IAO for H and a 2p-like IAO for N, directed along the NH$_2$-H axis. Such linear combinations have $\sigma$ and $\sigma^*$ character, as seen in Fig. \ref{fig:figure_3}.

\begin{figure}[h!]
\includegraphics[width=1.0\columnwidth]{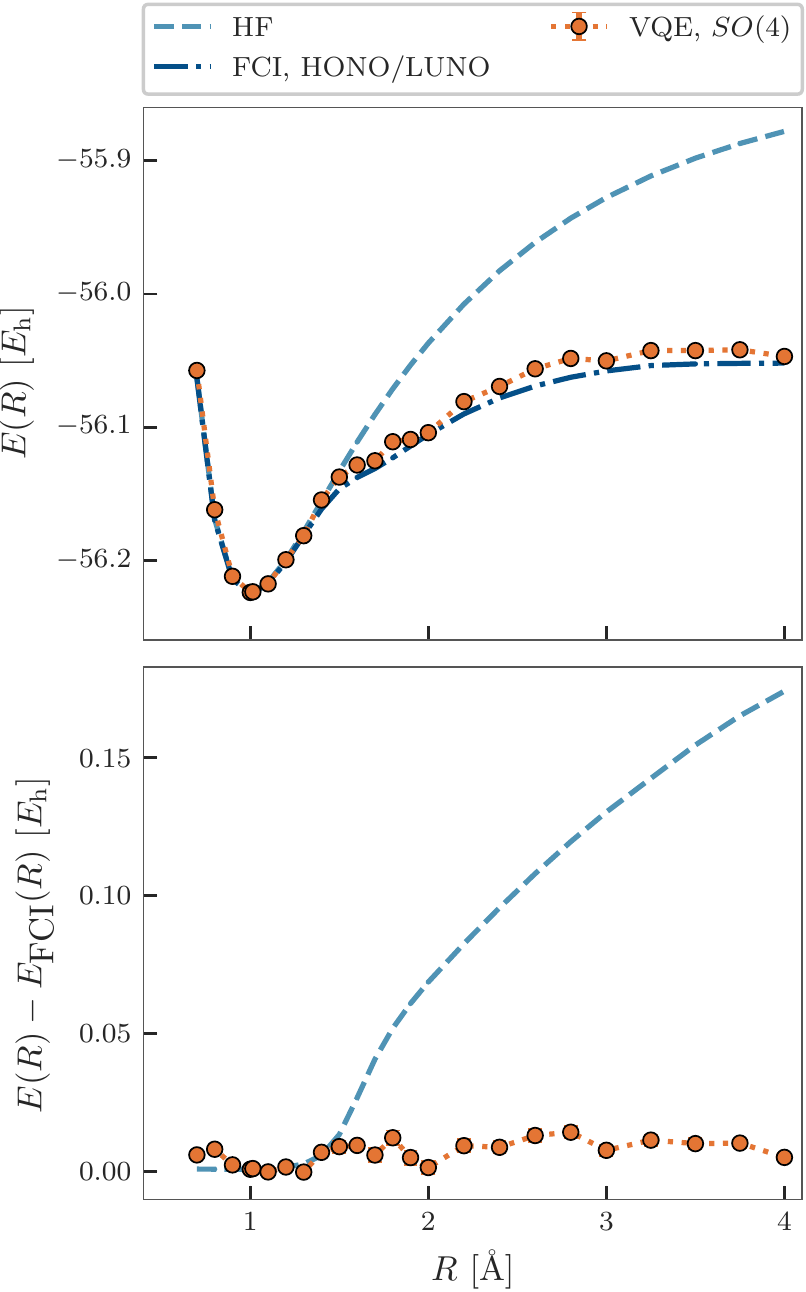} 
\caption{
Top: ground-state potential energy surface of NH$_3$ along the NH$_3 \to$ NH$_2$ + H reaction path, 
using a HONO/LUNO active space based on an MP2 calculation at IAO/aug-cc-pVQZ level, from RHF (dashed light blue line), FCI (blue dash-dotted line) and VQE with depth-1 SO(4) Ansatz (orange circles).
Bottom: deviation of RHF and VQE energies from FCI.
}
\label{fig:figure_4}
\end{figure}

The quantum circuit used to simulate the ground state of NH$_3$ is shown in Fig. \ref{fig:figure_3}.
Qubits are entangled through an SO(4) gate, parametrized leveraging the isomorphism between SO(4) and SU(2) $\times$ SU(2) \cite{vatan2004optimal,zulehner2019compiling}.
Parameters are optimized using a combination of analytical gradient evaluation \cite{parrish2019hybrid} and the gradient descent technique, as illustrated in Appendix \ref{app:opt}.
The VQE ground-state potential energy curve is shown in Fig. \ref{fig:figure_4}. 
As seen, VQE improves significantly over RHF in the active space, and yields results in qualitative agreement with FCI.

We emphasize that the use of the full-valence IAO basis, as in the H$_2$ application, is known to be reasonable from a chemical standpoint. 
Any other active space construction, as in the NH$_3$ application, needs to be supported on chemical grounds or assessed with numerical data.
To assess the accuracy of the HONO/LUNO active space and of our VQE simulations, in Table \ref{tab:table_2} we list 
active-space equilibrium bondlengths and binding energies, as well as the mean deviation
\begin{equation}
\Delta_{\mathrm{RHF,VQE}} = \sum_{i=1}^{N_R} \frac{ | E_{\mathrm{RHF,VQE}}(R_i) - E_{\mathrm{FCI}}(R_i) | }{N_R}
\end{equation}
of RHF and VQE results from FCI. Results indicate that VQE accurately reproduces FCI quantities within the HONO/LUNO active space.
\resub{Comparison between Tables \ref{tab:table_1} and \ref{tab:table_2}, on the other hand, 
indicates that the HONO/LUNO active space leads to a slightly shorter equilibrium bondlength than with the full IAO basis,
and to a less accurate binding energy. 
This is not unexpected, as the active space approximation affects the electronic structure to a varying extent along the dissociation profile.}

\begin{table}[h!]
\centering
\begin{tabular}{c|ccc}
\hline\hline
method                       & $\Delta E$ [$\mathrm{E_h}$] & $R_{eq}$ [$\mathrm{\AA}$] & $\Delta_{\mathrm{RHF,VQE}}$ [$\mathrm{E_h}$] \\
\hline
RHF                          & N/A      & 0.995(1)  & 0.060      \\
VQE, $\mathrm{SO}(4)$, $d=1$ & 0.165(3) & 1.007(3)  & 0.0032(17) \\
FCI                          & 0.180(2) & 0.9961(7) & N/A        \\
\hline\hline
\end{tabular}
\caption{N-H dissociation energy, NH$_3$ equilibrium bondlength and deviation from FCI, from RHF and VQE with depth-1 SO(4) Ansatz,
using a HONO/LUNO active space determined at IAO/aug-cc-pVQZ level.
}
\label{tab:table_2}
\end{table}

\subsubsection{Assessment of accuracy}
\label{sec:accuracy}

\begin{figure}[h!]
\includegraphics[width=1.0\columnwidth]{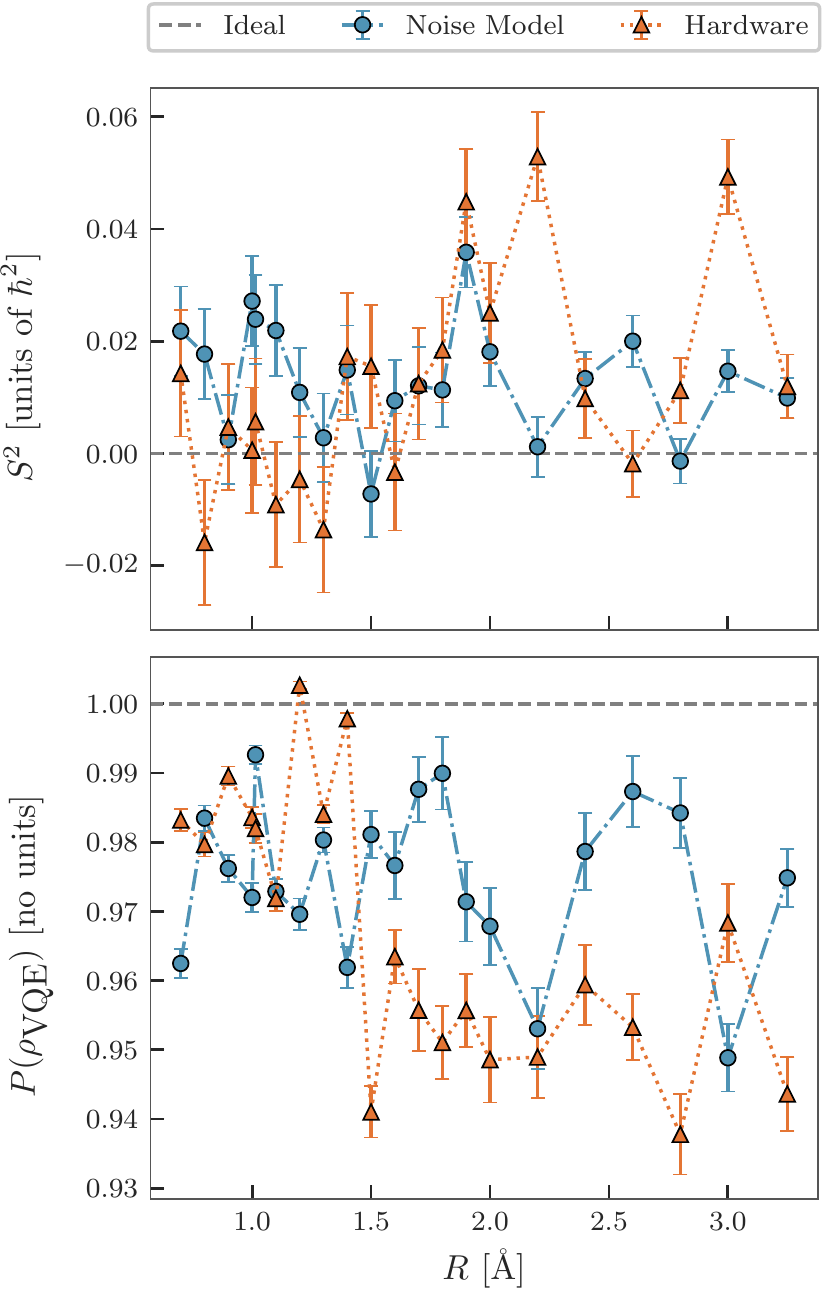} 
\caption{
Expectation values of the total spin operator (top) and purity of the VQE density operator (bottom)
as a function of the reaction coordinate $R$, evaluated over the VQE wavefunction with SO(4) Ansatz.
Orange and blue symbols denote hardware calculations carried out on \device{rome}, and on a a classical simulator with noise model from \device{manila}, respectively.
}
\label{fig:figure_5}
\end{figure}

Besides computing energies, it is important to gain as much insight as possible into the structure of the ground-state wavefunction. 
To achieve this goal, we compute the total spin operator $S^2$.
While such a quantity is a constant of motion, in simulations conducted on quantum hardware it may feature significant errors, due to decoherence.
As seen in Fig. \ref{fig:figure_5}, the VQE wavefunction is essentially in the singlet manifold $S^2=0$, but is not an eigenfunction of $S^2$.
Deviations from $S^2 = 0$ become slightly more intense for $R \geq 1.5 \mathrm{\AA}$, where the lowest-energy singlet and triplet states become nearly degenerate.
Further, we perform quantum state tomography (QST) \cite{longdellTOMO,steffenTOMO,bonkTOMO,maTOMO} 
over the VQE density operator $\rho_{\mathrm{VQE}}$,
\begin{equation}
\rho_{\mathrm{VQE}} = \sum_{\mu\nu=0}^3 
\frac{ \mbox{Tr}\left[ \rho_{\mathrm{VQE}} \left( \sigma^\mu \otimes \sigma^\nu \right) \right] }{4} 
\, 
\left( \sigma^\mu \otimes \sigma^\nu \right) \quad, \\
\end{equation}
where $\sigma^\mu$ is a Pauli operator with $\mu \in \{ id , x, y, z \}$ and $\sigma^{id}$ is the identity operator.
Using information from QST, we evaluate the purity
$ P\left(\rho_{\mathrm{VQE}}\right) = \mbox{Tr}\left( \rho_{\mathrm{VQE}}^2 \right) $
of the VQE density operator. 
$P(\rho) = 1$ if and only if $\rho$ is the projector $\rho = | \Psi \rangle \langle \Psi|$ onto a pure state $\Psi$. 
As seen in Fig. \ref{fig:figure_5}, for $R \geq 1.5 \mathrm{\AA}$ we observe $P(\rho) \simeq 0.955$. 
The observed decrease in purity signals decoherent interaction with the environment, that ultimately limits the accuracy of VQE simulations.

To elucidate the origin of the deviations from $S^2 = 0$ and purity $P[\rho]=1$, 
in Fig. \ref{fig:figure_5} we compute these quantities on a classical simulator with noise model from \device{manila}. 
While noise models capture decoherence only partially, we regard these data as an indication that the loss of accuracy seen here
is explained by a combination of well-understood \cite{aleksandrowicz2019qiskit} 
qubit decoherence (amplitude damping, dephasing errors), measurement, and gate error (coherent, incoherent) mechanisms.
In particular, since qubit decoherence and measurement errors affect these simulations uniformly across dissociation, 
the main source of error is represented by gates.

\begin{figure*}[t!]
\includegraphics[width=0.95\textwidth]{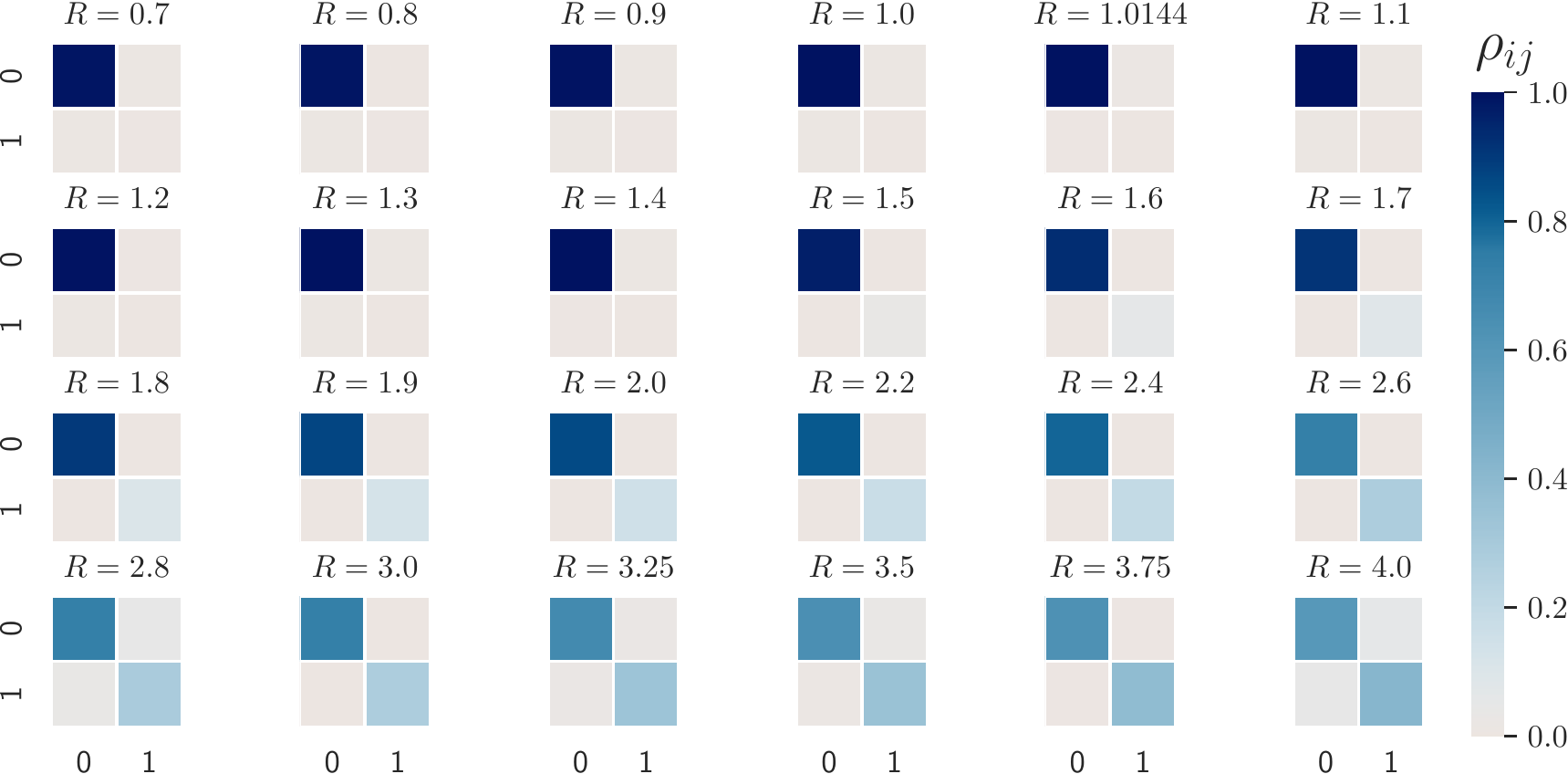} 
\caption{Spin-resolved one-body density matrix 
$\rho^{(\uparrow)}$ for NH$_3$ in a HONO-LUNO space, 
from VQE with SO(4) Ansatz, as a function of N-H distance 
(left to right, top to bottom).}
\label{fig:1rdm}
\end{figure*}

\begin{figure*}[t!]
\includegraphics[width=0.95\textwidth]{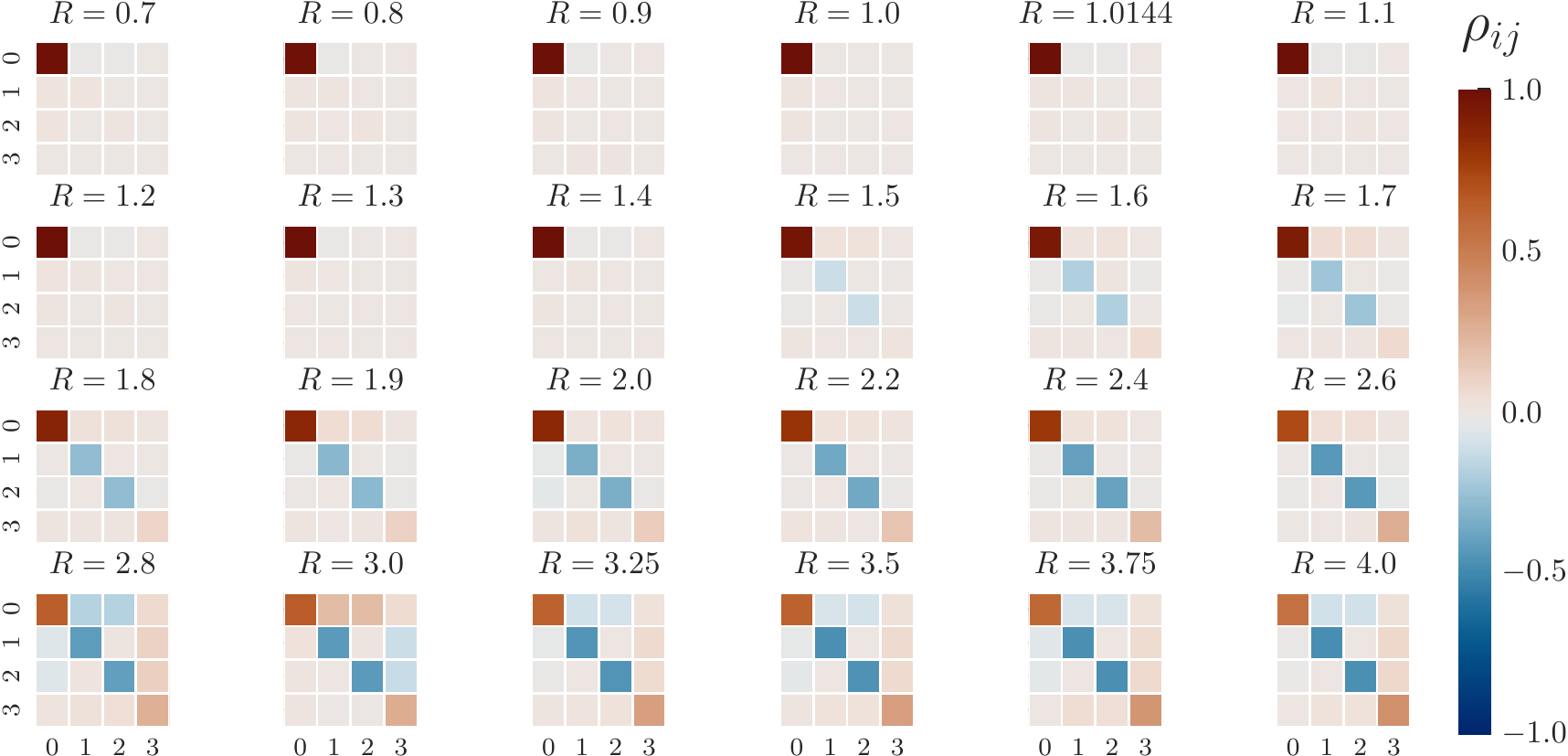} 
\caption{Spin-resolved two-body density matrix $\rho^{(\uparrow,\downarrow)}$ for NH$_3$ in 
a HONO-LUNO space, from VQE with SO(4) Ansatz, as a function of 
N-H distance (left to right, top to bottom). Numbers 0,1,2,3 denote
indices $(0,0)$, $(0,1)$, $(1,0)$, $(1,1)$ respectively.
}
\label{fig:2rdm}
\end{figure*}

\subsubsection{Density matrices}

The results shown in the previous Section \ref{sec:accuracy} are
mostly based on QST which, despite many recent theoretical and algorithmic improvements, 
remains an expensive operation with growing number $N_q$ of qubits \cite{leibfried1996experimental,poyatos1997complete,blume2013robust,merkel2013self,greenbaum2015introduction}.

An alternative way to obtain information about an electronic quantum state is provided by the one- and two-body density matrices, 
which can be obtained measuring up to $\mathcal{O}(N_q^5)$ qubit operators.

One- and two-body density matrices are shown in Figs. \ref{fig:1rdm} and \ref{fig:2rdm} respectively. The eigenvalues of the one-body density matrix evolve from $(1,0)$ to $(1/2,1/2)$ as $R$ increases, signaling that electrons become increasingly more entangled as the H atom separates from the NH$_2$ moiety. The same information is provided by the spin-resovled two-body density matrix 
$\rho^{(\uparrow,\downarrow)}_{prqs}$, which for small $R$ is peaked at $prqs=0000$, signaling that the ground state is approximately a single Slater determinant. As $R$ increases, 
$\rho^{(\uparrow,\downarrow)}_{0000} = \rho^{(\uparrow,\downarrow)}_{1111} \simeq 1/2$
and
$\rho^{(\uparrow,\downarrow)}_{0101} = \rho^{(\uparrow,\downarrow)}_{1010} \simeq -1/2$,
signaling that the ground state is a linear combination of two closed-shell singlet wavefunctions. 

\subsection{QITE hardware experiments}

\begin{figure}[h!]
\includegraphics[width=0.95\columnwidth]{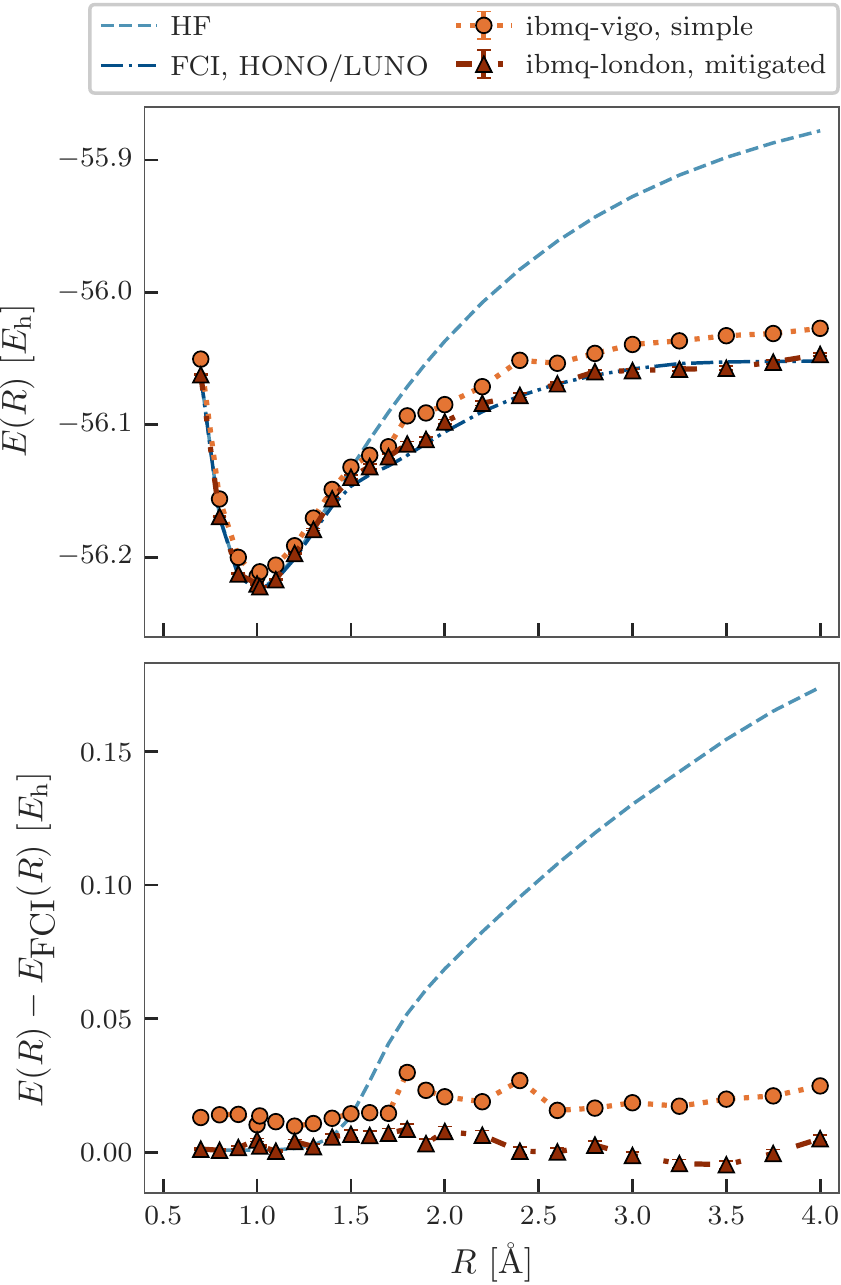} 
\caption{Ground-state potential energy surface of NH$_3$ along the NH$_3 \to$ NH$_2$ + H reaction path using quantum imaginary-time evolution, 
without (orange circles) and with (red triangles) readout error mitigation, on \device{vigo} and \device{london} respectively.}
\label{fig:qite}
\end{figure}

In Fig. \ref{fig:qite}, we further investigate the ground-state potential energy surface of NH$_3$ using the QITE method, using the 5-qubit \device{vigo} and \device{london} 
IBM Quantum hardware.
Details of QITE simulations, and especially simplifications made possible by the 2-qubit nature of the problem, are given in Appendix \ref{app:qite}.
In Fig. \ref{fig:qite} and Table \ref{tab:qite} we can appreciate the impact of readout error mitigation techniques \cite{kandala2019error,temme2017error,mcardle2019error,bravyi2020mitigating} on the accuracy of QITE, in terms of deviations from FCI as well as equilibrium bondlength and binding energy.
Readout error mitigation has more pronounced effect on results in the regime $R \geq 1.75$ where the electronic wavefunction starts acquiring multireference character and deviating appreciably from the Hartree-Fock state. 
Therefore, it does not affect the equilibrium bondlength within statistical uncertainties, whereas it affects the binding energy of the system by 
\resub{$\sim 0.015 \, \mathrm{E_h}$}.

\begin{table}[h!]
\centering
\begin{tabular}{c|ccc}
\hline\hline
method                       & $\Delta E$ [$\mathrm{E_h}$] & $R_{eq}$ [$\mathrm{\AA}$] & $\Delta_{\mathrm{RHF,QITE}}$ [$\mathrm{E_h}$] \\
\hline
RHF                          & N/A      & 0.995(1)  & 0.060      \\
QITE, no mitigation          & 0.195(4) &  1.002(9)  & 0.0169 \\
QITE, mitigation             & 0.178(2) &  0.993(7) &  0.0036 \\
FCI                          & 0.180(2) & 0.9961(7) & N/A        \\
\hline\hline
\end{tabular}
\caption{N-H dissociation energy, NH$_3$ equilibrium bondlength 
and deviation from FCI, from RHF and QITE with and without error mitigation,
using a HONO/LUNO active space determined at IAO/aug-cc-pVQZ level.
}
\label{tab:qite}
\end{table}

\section{Discussion}
\label{sec:conclusions}

In this work, we explored the use of intrinsic atomic orbitals in lieu of minimal basis sets of Gaussian orbitals in quantum simulations of molecular systems. Bases of IAOs have the same size of minimal bases, but offer more accurate estimates of energy differences and equilibrium geometries. IAOs arise from an exceptionally simple algebraic construction, require only mean-field calculations in larger basis sets to be defined, and draw a simple and effective connection between chemical concepts and numerical simulations.
As such, they are a compelling alternative to minimal basis sets in quantum simulations, along with other recently proposed approaches \cite{kottmann2020reducing,takeshita2020increasing}, until the progress of hardware and classical simulators of quantum computers will allow to routinely study larger basis sets from systematic sequences.

\resub{
The main limitation of IAOs is that electronic correlation is captured within a valence space. Therefore, perturbative or full inclusion of virtual orbitals is necessary to cover 
the dynamical correlations with methods like coupled cluster and multireference configuration interaction model, and very important to obtain quantitative agreement with 
experimental values, especially for sensitive quantities such as polarizabilities or thermochemical properties.
The connection between IAOs and larger bases can be leveraged to perturbatively include virtual orbitals beyond the IAO in the simulation, 
as we demonstrated here using a simplified implementation of VQSE for two-electron systems.}

We expect that the combination of intrinsic atomic orbitals, to partially overcome the limitations of minimal basis sets, and of density operators, to diagnose important properties of electronic wavefunctions, will prove useful tools in the simulation of chemical species by quantum algorithms on contemporary quantum devices.

\section*{Code availability}

The code used to generate the data presented in this study can be publicly accessed on GitHub at \cite{barison2020gihtub}.

\section*{Acknowledgments}

SB, DEG and MM acknowledge the Universit\`a degli Studi di Milano INDACO Platform and the IBM Research Cognitive Computing Cluster service respectively, for providing resources that have contributed to the results reported within this paper. SB acknowledges Sebastian Hassinger for help obtaining access to IBM Quantum hardware, and Jeffrey Cohn and Gavin Jones for helpful discussions. SB, MM and DEG acknowledge Gerald Knizia for helpful discussions. 




\newpage\hbox{}\thispagestyle{empty}\newpage

\appendix

\section{Comparison of IAO against other bases}
\label{app:iao}

\begin{figure}[h!]
\includegraphics[width=1.0\columnwidth]{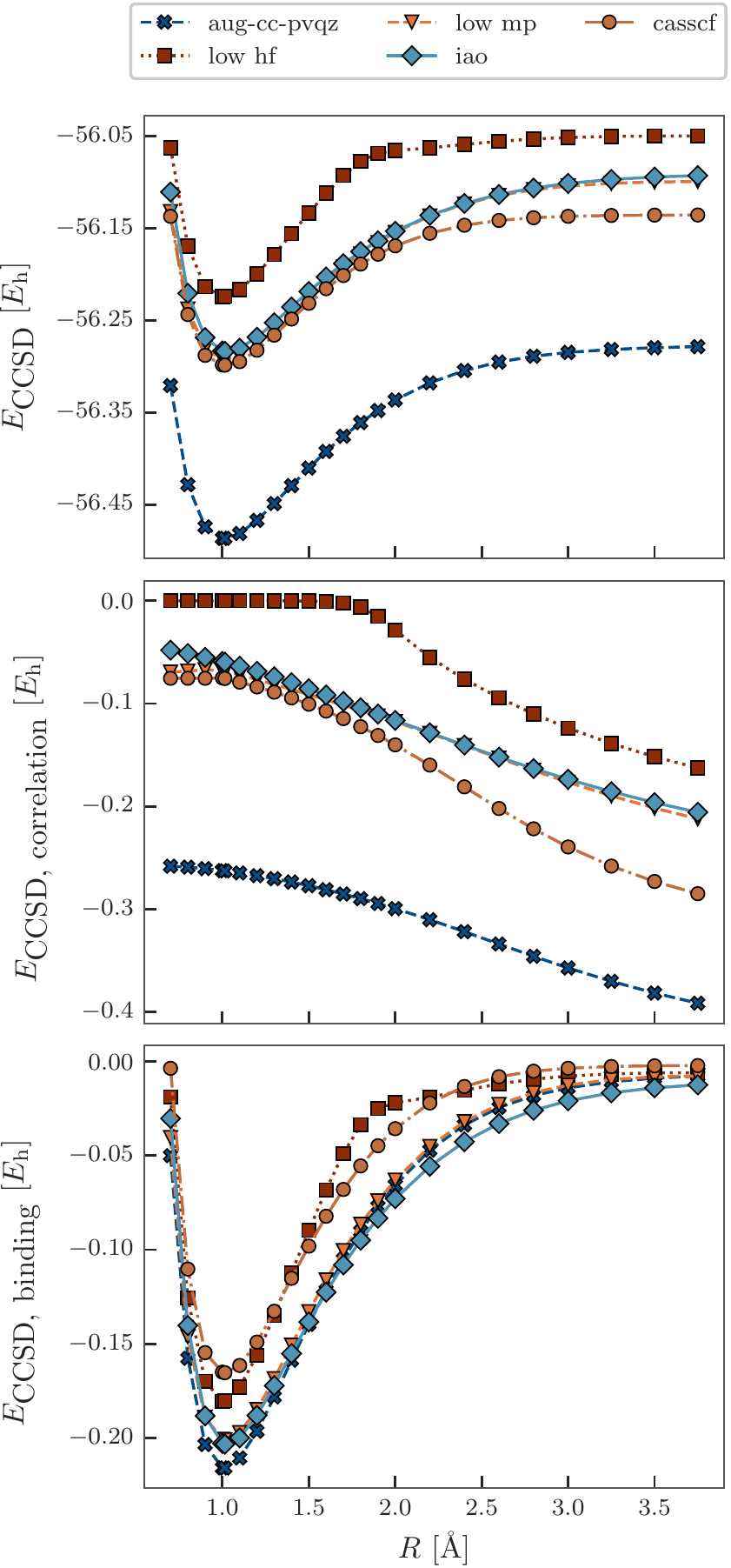} 
\caption{CCSD total (top), correlation (middle) and binding (bottom) energies
of NH$_3$ along the NH$_3$ $\to$ NH$_2$ + H dissociation path
from CCSD/aug-cc-pVQZ (blue crosses), low-energy Hartree-Fock orbitals 
(red squares), low-energy CASSCF orbitals (dark orange circles), high-occupancy
MP2 natural orbitals (orange triangles) and IAO/aug-cc-pVQZ (light blue diamonds).}
\label{fig:iao_rodeo}
\end{figure}

\begin{figure}[h!]
\includegraphics[width=1.0\columnwidth]{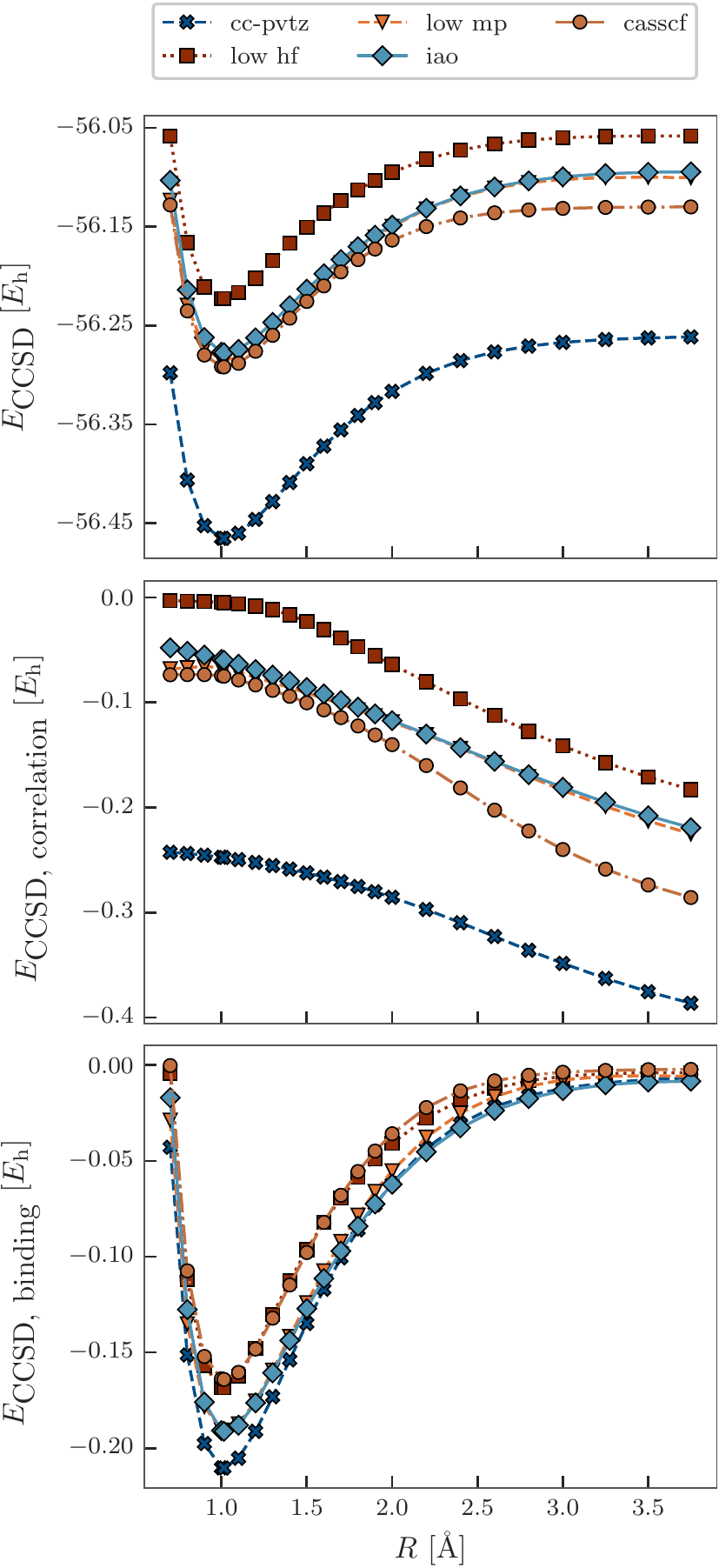} 
\caption{Same as Fig. \ref{fig:iao_rodeo} but for IAO/cc-pVTZ.}
\label{fig:iao_rodeo2}
\end{figure}

In this Section, we compare IAO potential energy curves along 
the NH$_3$ dissociation path, as well as binding energies and
equilibrium bondlengths, against those from active spaces of 
low-energy Hartree-Fock and CASSCF (complete active space 
self-consistent field) orbitals, and high-occupancy MP2 natural
orbitals.
Results are given in Figs. \ref{fig:iao_rodeo} and \ref{fig:iao_rodeo2}, 
using CCSD at aug-cc-pVQZ and cc-pVTZ level.

As seen, active spaces of low-energy Hartree-Fock give lower-accuracy 
total, correlation and binding energies than the other choices.
We reason that the worse performance of low-energy Hartree-Fock orbitals
is due to the inclusion of Rydberg, rather than anti-bonding, orbitals 
in the active space.
IAO performs similarly to high-occupancy MP2 natural orbitals, and
overall they give binding energies in better agreement with CCSD/aug-cc-pVQZ
than low-energy Hartree-Fock and CASSCF orbitals.

Compared against minimal bases, IAOs comprise higher-quality orbitals.
Compared against larger basis sets, the main benefit of IAOs is the reduced number 
of orbitals and qubits.
Numerical simulations with classical emulators employed
$(N_\alpha,N_\beta) = (4,4)$ and $(4,4)$ electrons, 
$N_{IAO} = 6$ and $7$ orbitals, and $N_{qubit} = 8$ and $11$ qubits
for water and ammonia respectively.

\section{Quantum Equation-of-Motion}
\label{app:qeom}
In this Appendix we turn our attention to electronic excited states, that we  investigate using the quantum equation-of-motion formalism.
The quantum Equation-of-Motion (qEOM) \cite{ollitrault2019quantum,gao2020applications,ollitrault2020hardware} is a technique for approximating excited states of quantum systems by applying suitable excitation operators to their ground state,
\begin{equation}
| \Psi_I \rangle = \hat{O}^\dagger_I | \Psi_0 \rangle \quad.
\end{equation}
In general, excitation operators are arbitrarily complicated many-body operators.
As in classical coupled-cluster calculations \cite{monkhorst1977calculation,stanton1993equation,krylov2006spin}, accurate approximations for selected excited states are obtained assuming that excitation operators are low-rank,
\begin{equation}
\begin{split}
\hat{O}^\dagger_I &= \sum_\mu X_{\mu I} \hat{E}_\mu - Y_{\mu I} \hat{E}^\dagger_\mu \;, \\
\hat{E}_\mu &\in \Big\{ \sum_\sigma \crt{a \sigma} \dst{i \sigma} , \sum_{\sigma\tau} \crt{a \sigma} \crt{b \tau} \dst{j \tau} \dst{i \sigma} \Big\}
\;,
\end{split}
\end{equation}
where indices $ij$ and $ab$ label occupied
and virtual orbitals in a mean-field reference state. The expansion coefficients are determined \cite{rowe1968equations,ollitrault2019quantum} solving a generalized eigenvalue equation of the form
\begin{equation}
\hspace{-0.2cm}
\left[
\begin{array}{cc}
M & Q \\
Q^* & M^* \\
\end{array}
\right]
\left[
\begin{array}{cc}
X_I \\
Y_I \\
\end{array}
\right]
=
\Delta E_I
\left[
\begin{array}{cc}
V & W \\
-W^* & -V^* \\
\end{array}
\right]
\left[
\begin{array}{cc}
X_I \\
Y_I \\
\end{array}
\right] \;,
\end{equation}
where matrix elements are defined as
\begin{equation}
\begin{split}
V_{\mu\nu} &= \langle \Psi | [\hat{E}^\dagger_\mu , \hat{E}_\nu ] | \Psi \rangle \\
M_{\mu\nu} &= \langle \Psi | [\hat{E}^\dagger_\mu , \hat{H} , \hat{E}_\nu ] | \Psi \rangle \\
W_{\mu\nu} &= - \langle \Psi | [\hat{E}^\dagger_\mu , \hat{E}^\dagger_\nu ] | \Psi \rangle \\
Q_{\mu\nu} &= - \langle \Psi | [\hat{E}^\dagger_\mu , \hat{H} , \hat{E}^\dagger_\nu ] | \Psi \rangle \\
\end{split}
\end{equation}
and triple commutators have the form 
\begin{equation}
[\hat{A} , \hat{B} , \hat{C} ] = 
\frac{
[[\hat{A} , \hat{B} ] , \hat{C} ] +
[\hat{A} , [\hat{B}  , \hat{C}] ] }
{2} \quad .
\end{equation}

\subsection{qEOM hardware experiments}

\begin{figure}[h!]
\includegraphics[width=1.0\columnwidth]{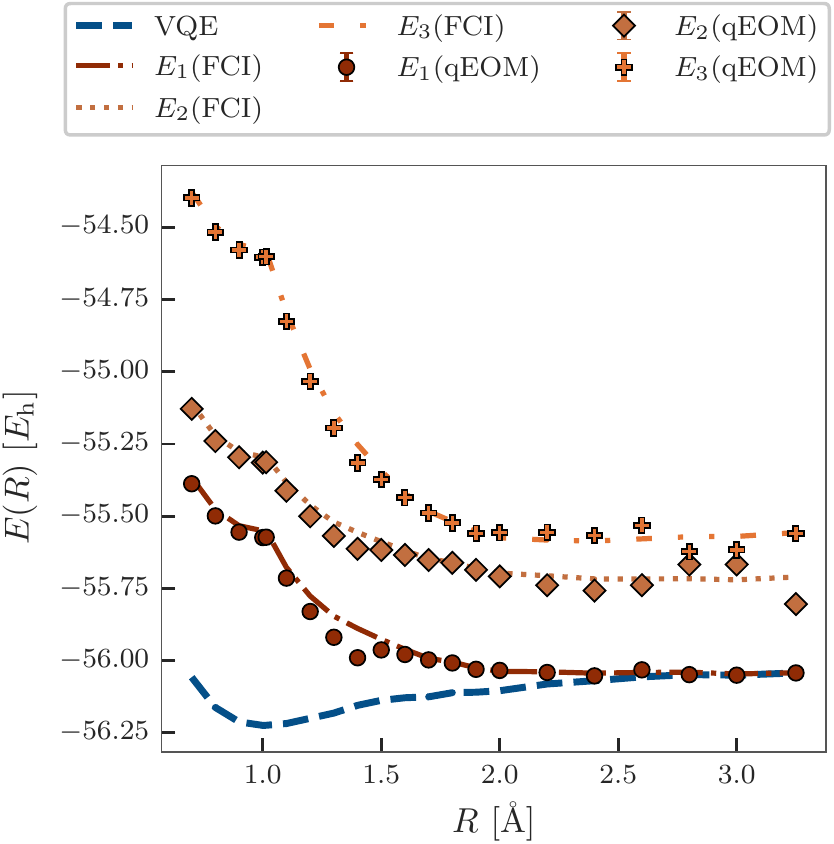} 
\caption{Excited-state energies (brown-red, dark-orange and orange for first, second and third excited state) of NH$_3$ along the NH$_3 \to$ NH$_2$ + H reaction path using FCI (lines) and qEOM (symbols), on the
{{\em{ibmq$\_$rome}}} IBM Quantum hardware. $E_{n}(M)$ indicates the $n$-th excited state obtained with method $M$. The blue dashed line indicates the energy calculated using the VQE on hardware.}
\label{fig:qeom}
\end{figure}

In Fig. \ref{fig:qeom} we show the qEOM energies of excited states in the HONO/LUNO subspace, using
{{\em{ibmq$\_$rome}}} with readout error mitigation. We mention that further mitigation of gate and 
readout error can be achieved by QST \cite{gao2020applications}, but for the purpose 
of the present work we elected to use the more standard readout error mitigation implemented in Qiskit and explained in Section \ref{sec:err_mitigation}.

The mean deviations between exact and computed excited-state energies is 0.019784, 0.027757 and 0.029781 for first, second and third excited state respectively. Of course, the use of a 2-orbital active space determined the ability to detect only a subset of excited states, that around the equilibrium geometry are significantly biased (discontinuities at $R \simeq 1$ \AA).
In the long $R$ limit, the ground and lowest excited state, of triplet character, become degenerate. Due to such degeneracy, the qEOM eigenvalue equation becomes ill-conditioned, as documented below, resulting in excited-state energies with lower accuracy than in the short $R$ regime.

\subsection{Details of qEOM simulations}

\begin{figure}[h!]
\includegraphics[width=1.0\columnwidth]{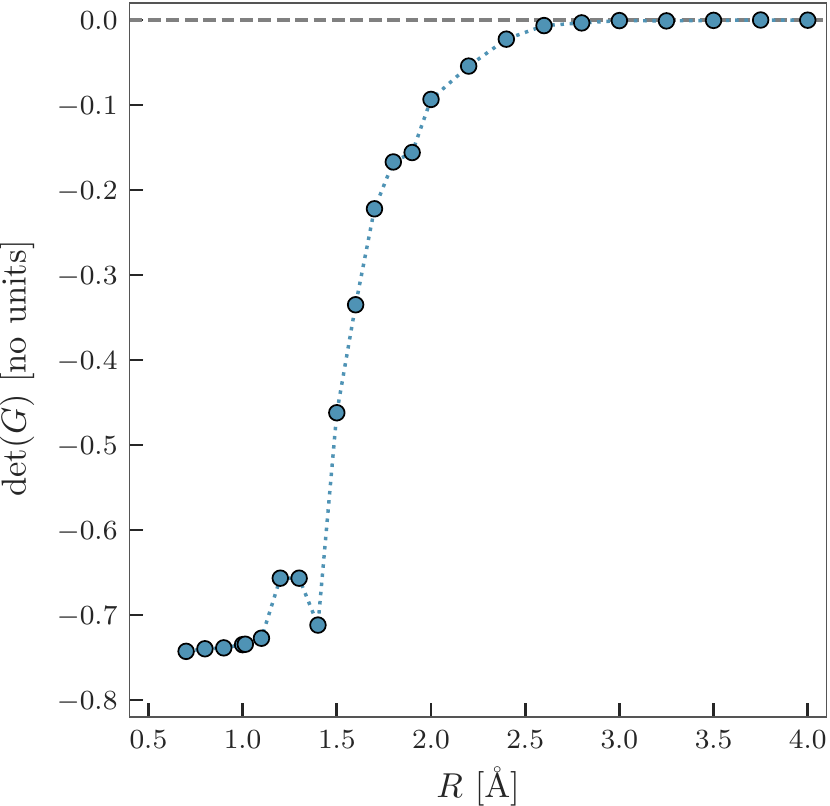} 
\caption{Determinant of the qEOM metric matrix G as a function of reaction coordinate, 
measured on {\em{ibmq$\_$rome}} (dotted, blue circles). The grey dashed line highlights $\mbox{det}(G) = 0$.}
\label{fig:det}
\end{figure}

Solving the qEOM equation $H u_i = \epsilon_i G u_i$, where we will call $H$ and $G$ 
the ``Hamiltonian" and ``metric" matrices respectively, requires the metric matrix $G$
to be numerically well-conditioned, and in particular to have 
$|\mbox{det}(G)| \gg 0$.
In Fig. \ref{fig:det}, we report the determinant $\mbox{det}(G)$ of the metric 
matrix as a function of reaction coordinate $R$ along the dissociation of ammonia.
As seen, for $R \geq 2.5$, the determinant approaches zero, signaling the incipient
degeneracy of singlet and triplet states. 

\section{Details of VQE simulations}
\label{app:vqe}

\subsection{Optimization on quantum hardware}
\label{app:opt}

The variational parameters $\theta$ are concentrated, for both 
the SO(4) and the Ry Ans\"{a}tze, in the angles of single-qubit
rotations. Given a unitary $\hat{U}(\theta)$ where a parameter 
$\theta_\mu$ appears in a single-qubit rotation only, it is 
known \cite{parrish2019hybrid} that
\begin{equation}
\begin{split}
\frac{\partial E}{\partial \theta_\mu}(\theta) &= E(\theta_+) - E(\theta_-)
\quad, \\
E(\theta_\pm) &= E\left( \dots \theta_\mu \pm \frac{\pi}{2} \dots \right)
\quad.
\end{split}
\end{equation}
Thus, the gradient of the VQE energy with SO(4) Ansatz can be 
computed analytically with 12 $n_{gates}$ energy measurements,
where $n_{gates}$ is the number of SO(4) gates in the circuit.

\begin{figure}[h!]
\includegraphics[width=1.0\columnwidth]{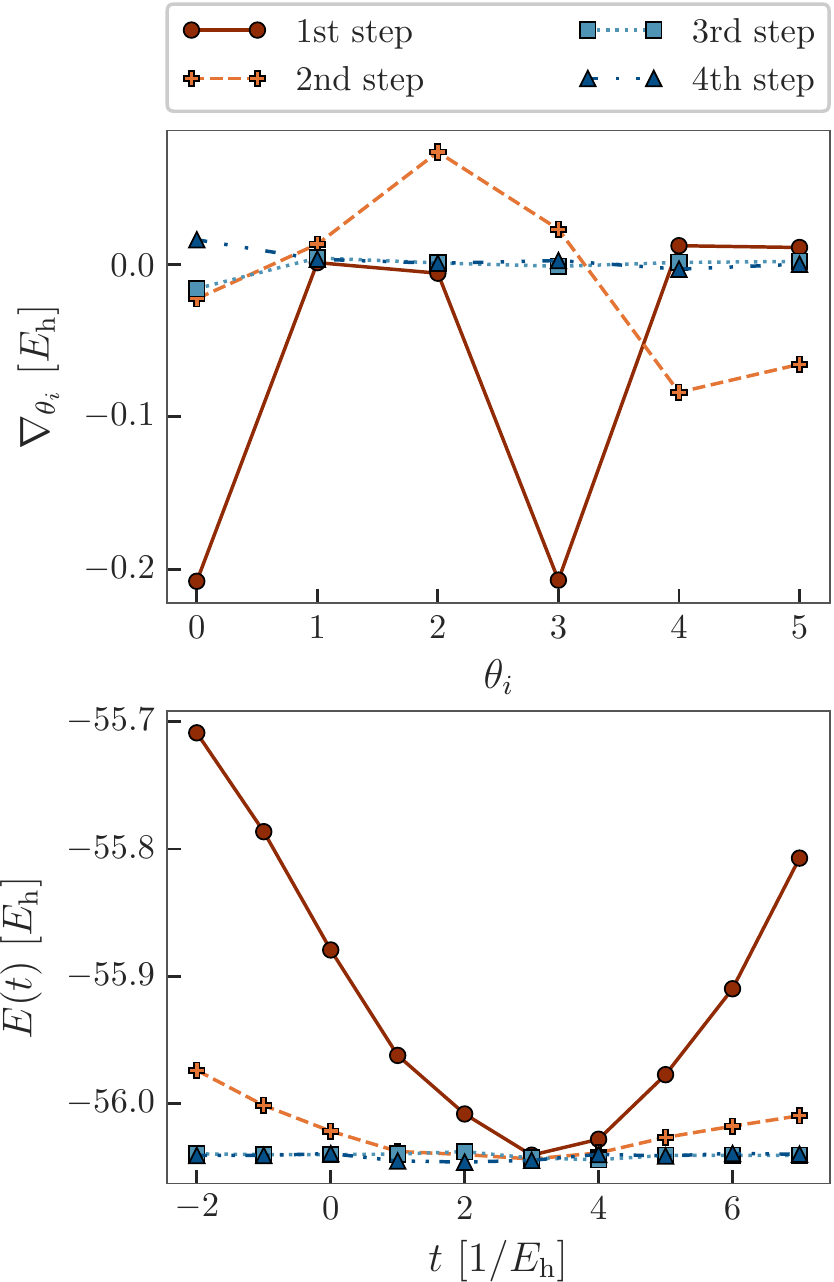} 
\caption{Analytical gradient evaluation (top) and line search (bottom)
in the gradient-descent optimization of VQE parameters with SO(4) Ansatz.}
\label{fig:steepest}
\end{figure}

In Fig. \ref{fig:steepest} we demonstrate SO(4) parameter optimization
by gradient descent at reaction coordinate $R=3.0$ \AA. In the 
gradient descent optimization scheme, parameters are initialized
from a configuration $\theta^{(0)}$, in our case $\theta^{(0)} = 0$
and, between iterations $i$ and $i+1$, are updated as
\begin{equation}
\begin{split}
\vec{\theta}^{(i+1)} &= \vec{\theta}^{(i)} - \lambda^* \, \vec{g}^{(i)}
\quad, \\
\vec{g}^{(i)} &= \vec{\nabla} E\Big( \vec{\theta}^{(i)} \Big) \quad, \\
\lambda^* &= \mbox{argmin}_{\lambda} \, E\Big(\vec{\theta}^{(i)} - \lambda \, \vec{g}^{(i)}\Big) \quad .
\end{split}
\end{equation}
The gradient $\vec{g}^{(i)}$ is computed analytically as detailed above.
The line search is performed manually at each iteration, and optimization
continues until convergence of the energy within statistical uncertainties.

\subsection{Fidelity between VQE and Hartree-Fock states}

\begin{figure}[h!]
\vspace{0.3cm}
\includegraphics[width=1.0\columnwidth]{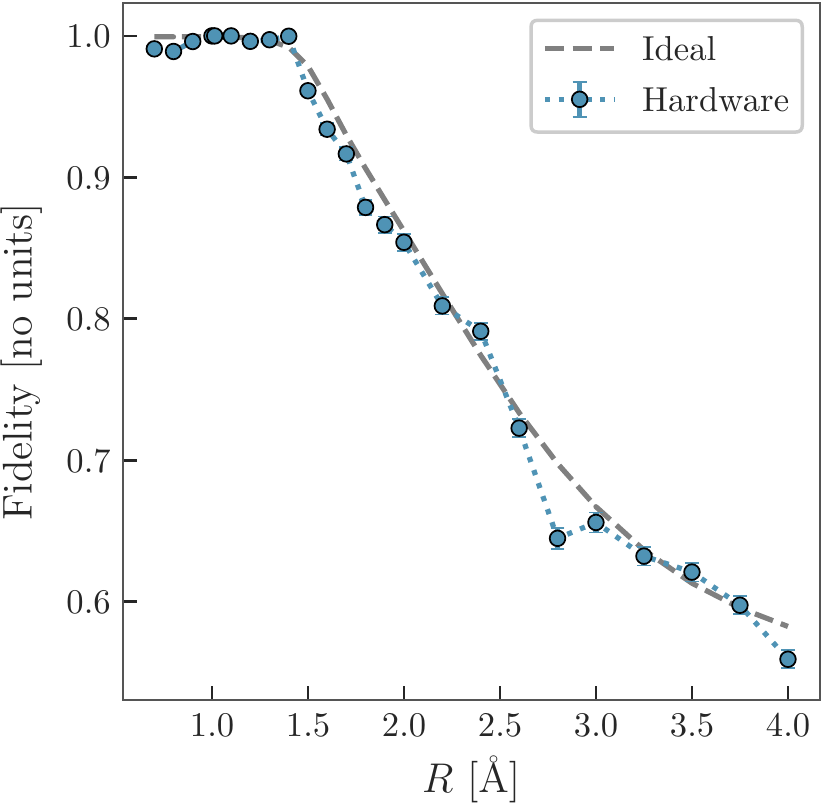} 
\caption{Fidelity between the VQE density operator and the projector onto the RHF state as a function of reaction coordinate. The grey dotted line indicates the ideal result of a statevector simulation. The blue circles are the results of the experiment on \device{rome}.}
\label{fig:fidelity}
\end{figure}

To gain further insight in the structure of the wavefunction, we used 
information from the measurement of density matrices to evaluate the fidelity
\begin{equation}
F\big[\rho_{\mathrm{VQE}},| \Psi_{\mathrm{RHF}} \rangle \langle \Psi_{\mathrm{RHF}}|\big] = \langle \Psi_{\mathrm{RHF}} | \rho_{\mathrm{VQE}} | \Psi_{\mathrm{RHF}} \rangle
\end{equation}
between the VQE density operator and the projector onto the RHF state, shown in Fig. \ref{fig:fidelity} as a function of reaction coordinate. Interestingly, both deviations from $S^2 = 0$ and decrease in purity are concomitant with the decrease in fidelity between VQE density operator and RHF, starting at $R \geq 1.5 \, \AA$,
and signalling acquisition of multireference character by the VQE density operator.

\section{Details of QITE simulations}
\label{app:qite}

In the QITE simulations performed here, we use a time step \resub{$\Delta \tau = 0.5 \mathrm{E_h}$} and a total projection time \resub{$\beta = 7.0 \mathrm{E_h}$}. Since the Hamiltonian $\hat{H}$ is a two-qubit operator, we perform imaginary-time evolution under the operator $\hat{H}$ without Trotter-Suzuki 
or similar approximations. Imaginary-time evolution is reproduced by two-qubit unitaries, ensuring that computed quantities agree with 

To keep the circuit depth and the number of CNOT gates in the QITE circuit constant as $\beta$ increases, we rely on a KAK decomposition 
\cite{kraus_2001}: the QITE unitary for $n$ time steps, $U_n$, is computed on the classical computer and reduced to a quantum circuit comprising 2 CNOT gates \cite{kraus_2001}.

Such a technique, used for example in the context of spin simulations \cite{francis2020magnon,sun2020quantum}, is specifically designed 
for two-qubit systems. Research to generalize these approximations to more general situations is underway.

\section{Variational quantum subspace expansion}
\label{app:vqse}

In this Appendix, we review briefly VQSE technique, proposed by Takeshita et al \cite{takeshita2020increasing}, and calculate the explicit expression of matrix elements for our problem.
We  show that all the matrices can be evaluated using data from a quantum hardware.

As indicated in the main text, the starting point of VQSE is a reference function $\Psi_0$ constructed in a set of active orbitals from a large basis.
Here, active-space orbitals are linear combinations of IAOs, denoted with lowercase letters, $p \in A$.
Uppercase letters $P \in B_1$ denote orthonormal orbitals in the basis used to construct IAOs.

Next, VQSE introduces a set of expansion operators.
Here, we choose

\begin{equation}
\begin{split}
| \Psi \rangle 
&= \left[ \alpha + \beta_{P r} \crt{P \sigma} \dst{r \sigma} + \gamma_{TuVw} \crt{T \sigma} \crt{V \tau} \dst{w \tau} \dst{u \sigma} \right]  | \Psi_0 \rangle \\
&= \left[ \alpha + \beta_{P r} E_{Pr} + \gamma_{TuVw} E_{TuVw} \right] | \Psi_0 \rangle \, .
\end{split}
\end{equation}

Electrons are excited from active to generic orbitals, excitation operators are summed over spin polarizations $\sigma,\tau$
and Einstein's summation convention is used.
Note that the reference wavefunction has no components outside the active space $A$, 
and therefore contraction over orbitals outside $A$ can be computed analytically using Wick's theorem.

The amplitudes $v = \left( \, \alpha \,\,\, \beta \, \, \, \gamma \, \right)^T$
are real-valued, and determined by solving a generalized eigenvalue equation $H v = E S v$.

The matrices $H$ and $S$ are defined by the bilinear forms
\begin{widetext}
\begin{equation}
\label{eq:vqse}
\begin{split}
\langle \Psi | \Psi \rangle 
&= 
( \alpha , \beta_{Qs} , \gamma_{XyZa} ) 
\left(
\begin{array}{ccc}
1 & \langle \hat{E}_{Pr} \rangle & \langle \hat{E}_{TuVw} \rangle \\
\langle \hat{E}_{sQ} \rangle & \langle \hat{E}_{sQ} E_{Pr} \rangle & \langle \hat{E}_{sQ} \hat{E}_{TuVw} \rangle \\
\langle \hat{E}_{yXaZ} \rangle & \langle \hat{E}_{yXaZ} E_{Pr} \rangle & \langle \hat{E}_{yXaZ} \hat{E}_{TuVw} \rangle \\
\end{array}
\right) 
\left(
\begin{array}{c}
\alpha \\
\beta_{Pr} \\
\gamma_{TuVw} \\
\end{array}
\right) \\
\langle \Psi | \hat{H} | \Psi \rangle 
&= 
( \alpha , \beta_{Qs} , \gamma_{XyZa} ) 
\left(
\begin{array}{ccc}
\langle \hat{H} \rangle & \langle \hat{H} \hat{E}_{Pr} \rangle & \langle \hat{H} \hat{E}_{TuVw} \rangle \\
\langle \hat{E}_{sQ} \hat{H} \rangle & \langle \hat{E}_{sQ} \hat{H} \hat{E}_{Pr} \rangle & \langle \hat{E}_{sQ} \hat{H} \hat{E}_{TuVw} \rangle \\
\langle \hat{E}_{yXaZ} \hat{H} \rangle & \langle \hat{E}_{yXaZ} \hat{H} \hat{E}_{pr} \rangle & \langle \hat{E}_{yXaZ} \hat{H} \hat{E}_{TuVw} \rangle \\
\end{array}
\right) 
\left(
\begin{array}{c}
\alpha \\
\beta_{Pr} \\
\gamma_{TuVw} \\
\end{array}
\right) \\
\end{split}
\end{equation}
\end{widetext}
Here, $\hat{H} = T_{EFGH} \crt{E \sigma} \crt{G \tau} \dst{H \tau} \dst{F \sigma}$ denotes the Hamiltonian, written compactly as a two-body operator,
and angular brackets denote expectation values over the reference state. 

In the present work, we focused on two-electron problems, where the bilinear forms in Eq.~\eqref{eq:vqse} are defined by the full-basis one- and two-body density matrices, 
which in turn can be trivially computed given their active-space counterparts \cite{takeshita2020increasing}.

A lengthy but straightforward calculation, based on Wick's theorem, shows that the bilinear forms in Eq.~\eqref{eq:vqse} are given by
\begin{widetext}
\begin{equation}
\label{eq:eqn_forms}
\begin{split}
\langle \hat{E}_{Pr} \rangle &= \rho_{Pr} \\
\langle \hat{E}_{TuVw} \rangle &= \rho_{TuVw} \\
\langle \hat{E}_{sQ} \rangle &= \rho_{sQ} \\
\langle \hat{E}_{sQ} \hat{E}_{Pr} \rangle &= \delta_{PQ} \rho_{sr} + \rho_{PrsQ} \\
\\
\langle \hat{E}_{sQ} \hat{E}_{TuVw} \rangle &= \delta_{QT} \rho_{suVw} - \delta_{QV} \rho_{suTw} \\
\\
\langle \hat{E}_{yXaZ} \rangle &= \rho_{aZyX} \\
\langle \hat{E}_{yXaZ} \hat{E}_{Pr} \rangle &= \delta_{XP} D_{aZyr} - \delta_{ZP} D_{yraX} \\
\\
\langle \hat{E}_{yXaZ} \hat{E}_{TuVw} \rangle &= \left[ \delta_{XT} \delta_{VZ} - \delta_{ZT} \delta_{XV} \right] D_{awyu} \\
\\
\end{split}
\hspace{1cm}
\begin{split}
\langle \hat{H} \hat{E}_{Pr} \rangle &= T_{EPGH} \; \rho_{ErGH} + T_{EFGP} \; \rho_{EFGr} \\
\langle \hat{H} \hat{E}_{TuVw} \rangle &= T_{EVGT} \; \rho_{EwGu} + T_{ETGV} \; \rho_{EuGw} \\
\langle \hat{E}_{sQ} \hat{H} \rangle &= T_{QFGH} \; \rho_{GHsF} + T_{EFGH} \; \rho_{EFsH}  \\
\langle \hat{E}_{sQ} \hat{H} \hat{E}_{Pr} \rangle &= T_{QFGP} \; \rho_{GrsF} + T_{QPGH} \; \rho_{GHsr} \\ &+ T_{EFQP} \; \rho_{QFsr} + T_{EPQH} \; \rho_{ErsH} \\
\langle \hat{E}_{sQ} \hat{H} \hat{E}_{TuVw} \rangle &= T_{QTGV} \; \rho_{Gwsu} + T_{EVQT} \; \rho_{Ewsu} \\ &+ T_{QVGT} \; \rho_{Gusw} + T_{ETQV} \; \rho_{Eusw} \\
\langle \hat{E}_{yXaZ} \hat{H} \rangle &= T_{XFZH} \; \rho_{aHyF} + T_{ZFXH} \; \rho_{aFyH} \\
\langle \hat{E}_{yXaZ} \hat{H} \hat{E}_{Pr} \rangle &= T_{XFZP} \; \rho_{aryF} + T_{ZFXP} \; \rho_{aFyr} \\ &+ T_{ZPXH} \; \rho_{aryH} + T_{XPZH} \; \rho_{aHyr} \\
\langle \hat{E}_{yXaZ} \hat{H} \hat{E}_{TuVw} \rangle &= \left[ T_{ZVXT} + T_{XTZV} \right] \; \rho_{awyu} \\ &+ \left[ T_{ZTXV} + T_{XVZT} \right] \; \rho_{auyw} \\
\end{split}
\end{equation}
\end{widetext}
Importantly, the simplifications in Eq.~\eqref{eq:eqn_forms} hold for two-electron active spaces.
Otherwise, three- and four-body active-space density matrices are required by VQSE,
leading to an $\mathcal{O}(N_o^8)$  computational cost in the number $N_o$ of active orbitals.
Active-space density matrices were computed with the techniques seen in the main text and embedded into their full-basis counterparts.

\end{document}